%% file: jgl.tex
\documentclass[runningheads,fleqn]{svmult}

\usepackage{makeidx}   
\usepackage{graphicx}  
\usepackage{subeqnar}  
\usepackage{multicol}  
\usepackage{cropmark} 
\usepackage{physmult}  
\makeindex             

\def\Journal#1#2#3#4{{#1} {\bf #2}, #3 (#4)}

\def\PLB{{\sl Phys. Lett.} {\bf B}}
\def\PRL{\sl Phys. Rev. Lett.}
\def\PRD{{\sl Phys. Rev.} {\bf D}}

\begin{document}

\title{The Atmospheric Neutrino Anomaly:\\
Muon Neutrino Disappearance}

\toctitle{Atmospheric Neutrino Anomaly}

\titlerunning{Atmospheric Neutrino Anomaly}

\author{John G. Learned}

\authorrunning{John G. Learned}

\maketitle              

\section{Introduction}

With the 1998 announcement of new evidence for muon neutrino
disappearance observed by the Super-Kamiokande experiment, the more
than a decade old atmospheric neutrino anomaly moved from a possible
indication for neutrino oscillations to an almost inescapable
implication.  In this chapter the evidence is reviewed, and
indications are presented that the oscillations are probably between
muon and tau neutrinos with maximal mixing.  Implications and future
directions are discussed.

The understanding of this phenomenon is now dominated by the data
announced by the Super-Kamiokande Collaboration at Neutrino98, of
which group the present author is a member.  Much of this report
dwells upon those results and updates to them, and so credit for this
work is due to the whole Collaboration, listed in the Appendix, who
have labored hard to bring this experiment to fruition and who have
been ably led by Prof. Yoji Totsuka of the University of Tokyo.  That
said, this report presents personal recollections and opinions of the
author, particularly in matters of the previous history,
interpretation of the present situation, and future prospects for this
line of research.

The phenomenon of neutrino oscillations is discussed in several other
Chapters of this volume (II and IX in particular), and to those the
reader is directed for derivation of the expressions utilized in this
Chapter and for understanding of the origin and implications of
neutrino oscillations generally.  Model building and implications in
astrophysics and cosmology are likewise treated elsewhere, while in
the following we focus narrowly upon the atmospheric neutrino anomaly,
its experimental explication in terms of muon neutrino oscillations
with tau neutrinos, and the implications of those results.

	\subsection{Atmospheric Neutrinos}

\begin{figure}
\includegraphics[width=.4\textwidth]{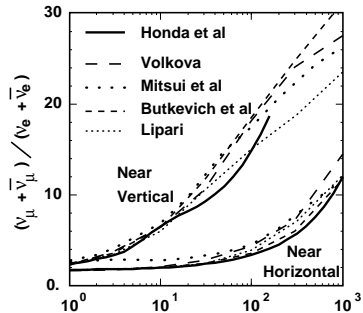}
\caption{The calculated ratio of the fluxes of atmospheric muon
neutrinos to electron neutrinos, versus neutrino energy.
Figure from Honda\cite{Honda_95}.}
\label{fig:mutoe}
\end{figure}

The neutrinos under discussion in this chapter arise from the decay
of pions and other mesons, and muons, which are produced in the
earth's atmosphere\cite{Gaisser_90_t,Berezinsky_90}.  The atmosphere
is being constantly bombarded with cosmic rays, which consist mostly
of protons, but also heavy nuclei and electrons, and even neutral
particles.  The earth's magnetic field, plus other magnetic fields
cut off the lower energy particles from the sun and more distant
sources, so that the mean incoming kinetic energy is around $1~GeV$.
Cosmic rays with lower energies do not cause effects which we can
directly detect on earth or underground.  Particles with energies in the
multi-GeV range make showers in the roughly ten-interaction-length-thick
(vertical column density) atmosphere.  Cosmic ray collisions
with air nuclei produce pions and other particles in abundance,
which themselves further interact or decay.  This competition
between interaction and decay leads to a steeper spectrum for the
decay products.  At energies below several GeV the muons produced in
charged pion decay, themselves decay:

\begin{equation}
\pi^+ \rightarrow \mu^+ \ \nu_{\mu}\\
\pi^- \rightarrow \mu^- \ \bar{\nu}_{\mu}\\
\mu^+ \rightarrow e^+ \ \nu_e \ {\bar\nu}_{\mu}\\
\mu^- \rightarrow e^- \ {\bar\nu}_e \ \nu_{\mu},
\end{equation}

with decay lengths of $L_{\pi^\pm} = 0.056~km\times E_\pi/GeV$ and
$L_\mu = 6.23~km \times E_\mu/GeV$.  Typical pion interactions lengths
(roughly $150~gm/cm^2$) are on the order of a few $km$ depending upon
altitude, angle and energy, while muons generally come to rest before
decaying or being absorbed.  Moreover (crucially and often ignored)
the energy sharing in the decays is such that the resulting neutrinos
are also of nearly equal energy.  These decay kinematics are of course
well known, so the ratio of muon neutrinos to electron neutrinos can
be calculated with rather good accuracy, about 5\%, almost
independently of the cosmic ray spectrum\cite{Stanev_99,Volkova_80},
as illustrated in Figure
\ref{fig:mutoe}.

Precise neutrino flux calculations (few percent) from man-made sources
are difficult if not impossible, as indicated in other chapters in
this book. The problem is even more difficult for the atmospheric
neutrinos, since the absolute magnitude of the incoming cosmic ray
flux is not well known, being uncertain at present to perhaps
25\%.\cite{Stanev_99}.  The calculations of the atmospheric neutrino
flux in the few $GeV$ energy range require not only the input cosmic
ray flux, with appropriate modulation to account for solar cycle
variation and geomagnetic field, but also details of nucleon-nucleon
and meson-nucleon interactions, not all of which have been well
measured.  The neutrino flux calculations also lead to {\it muon} flux
predictions, and these can be (and have been) compared to data, though
the appropriate data for low energy muons at high altitude, as
recorded in balloon measurements, is sparse and imprecise.  Typically
these calculations incorporate the approximation that the incoming
cosmic rays, the secondaries and even the neutrinos all travel in the
same direction.  This is no doubt not a serious compromise in the few
$GeV$ energy range, but has some effects in the energy range of a few
hundred $MeV$.  At this time new calculations are in
progress\cite{newfluxes}, but the computer time required to do the
full simulation is still a limiting factor and the job has yet to be
done definitively.

\begin{figure}
\includegraphics[width=.8\textwidth]{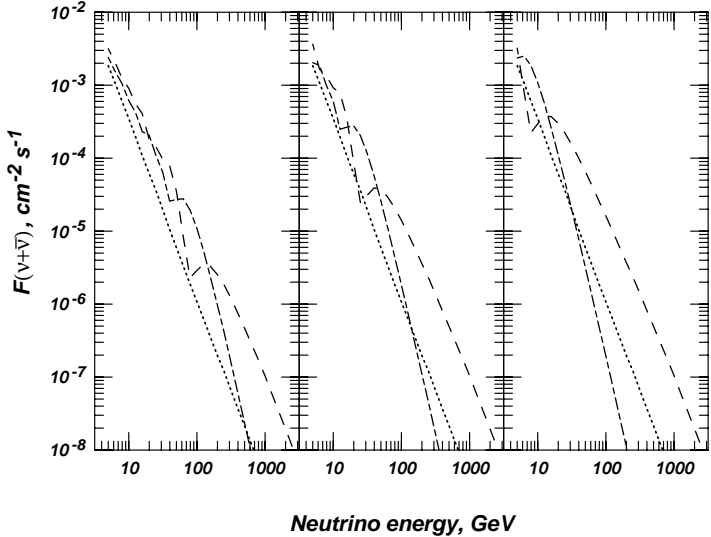}
\caption{Neutrino fluxes of all three flavors (dots: $\nu_e$,
dashes: $\nu_{\mu})$, long-short dashes $\nu_{\tau}$.) in the
presence of $\nu_{\mu} \rightarrow \nu_{\tau}$ oscillations with
maximal mixing and $\Delta m^2~=~10^{-2}, 10^{-2.5}, 10^{-3}~eV^2$,
from left to right.  Figure from Stanev\cite{Stanev_99}.}
\label{fig:atmnuspect}
\end{figure}

Several features of the neutrino flux are worth mentioning.  The muon
neutrino flux can be approximated as a power law with spectral index
$\gamma \simeq -3.7$\cite{Honda_95} for energies between about $10~GeV$ and
$100~TeV$.  The electron neutrino (and anti-neutrino) fluxes which
largely arise from muon decay, fall off more swiftly above several GeV,
with strong angle dependence. As illustrated in Figure
\ref{fig:mutoe} the $\nu_{\mu}$ to $\nu_e$ flux ratio falls to a few
percent at the higher energies, where the $\nu_e$'s are mostly
produced in kaon decay, $K^+ \rightarrow \pi^o e^+ \nu_e$ (4.82\% BR).
See Figure \ref{fig:atmnuspect} for atmospheric neutrino spectra of
all three flavors, as expected under several assumptions of
oscillations (two-flavor oscillations only).

There is a significant zenith angle variation in the atmospheric
neutrino flux, more prominent at higher energies, called the ``secant
theta'' effect.  This is simply due to the fact that those pions and
muons which are produced by incoming cosmic rays with trajectories
nearly tangential to the earth have more flight time in less dense
atmosphere, and hence more chance to decay.  Thus there is a peak,
becoming is more prominent at higher energies, near the horizontal
arrival direction in the atmospheric neutrino angular distribution.
This peak is symmetric about the horizon for any location except at
the lowest neutrino energies, below around $400~MeV$, where
geomagnetic effects spoil the symmetry somewhat.

The atmospheric neutrino energies practically accessible in
underground experiments range from the few tens of MeV to the 1 TeV
range, and the flight distances from roughly $20~km$ for down-coming
neutrinos to $13,000~km$ for those traversing the earth from the far
side.  The neutrino cross section is sufficiently small that there
should be negligible attenuation of these neutrinos: the attenuation
is roughly  $2.4 \times 10^{-5} E_{\nu}/GeV$ for neutrinos
traversing the earth's core, and thus negligible for any energies
below about $100~TeV$.  In consequence of the large dynamic range in
both energy and flight distance, the atmospheric neutrinos are
potentially sensitive to oscillations over a range of mass squared
differences from about $10^{-4}$ to $10^{-1}~eV^2$, as is discussed
below.

        \subsection{Initial Indications}

We will not dwell upon the past history\cite{Riordan}, but note that
the atmospheric neutrino anomaly has been around for some time,
roughly since the mid-1980's.  Indeed the first notice of something
peculiar in the atmospheric neutrino data stems from the 1960's when
the seminal underground experiments in South Africa\cite{CWI} and
South India\cite{KGF} first detected the natural neutrinos and
observed somewhat of an absolute rate deficit, but not convincingly as
the flux predictions were rough and the statistics small.

A new round of instruments were built beginning in the late 1970's to
search for nucleon decay as predicted by the (soon to be discarded)
SU(5) unification model.  The problem with atmospheric neutrinos, a
background to nucleon decay searches, became serious after the
activation of the first large water underground Cherenkov detector,
the IMB experiment, and by 1983 the realization that the number of
events containing muon decays was less than expected\cite{imb_anom}.
Soon this was confirmed by the second large water detector, the
Kamioka experiment\cite{kam_anom}, which group extended the results
with good particle identification giving a redundant measure of the
relative muon deficit (as also did IMB).  Some members of the
IMB\cite{imb} and the Kamioka\cite{jgl_kam} groups began to suggest, at
least in private, that oscillations were the cause of the deficit, but
that conclusion was not widely taken seriously for nearly ten
years. Indeed the first published interpretations of the anomaly as
due to oscillations were largely from outside the experimental
groups\cite{osc_hyp}.  To be fair though, it seems to be the Kamioka
group who first seriously believed that the anomaly was due to
oscillations and not simply a detector or background problem.

The deficit is characterized usually as an $R$ value, the double ratio
of muon to electron neutrinos, observed to expected. The effect was
large, the observed $R$ at about 2/3 of the expected value.  With the
initial evidence, the oscillations could have been from muon neutrinos
to others (eg. $\nu_{\tau}$ or a new neutrino) or between the muon and
electron neutrinos themselves.  It was the ratio that was in deficit:
one could not be sure whether there was an excess of electron
neutrinos, a deficit of muon neutrinos, or some of both.  This led to
suggestions of other possible ``physics'' causes, such as nucleon
decay favoring electron modes (since the anomaly was not initially
detected above the nucleon mass), or an excess of extraterrestrial
electron neutrinos.  See Table \ref{tab:hyp} for a graphical summary
of the situation.  There were also suggestions of systematic problems,
such as problems in muon identification, something wrong with flux
calculations or neutrino interaction cross sections, entering
backgrounds, or generic problems with the water Cherenkov detectors.

Over the intervening years between the emergence of this ``atmospheric
neutrino anomaly'', as it became known, and the 1998 SuperK
announcement, a great deal of effort went into study of these possible
systematic causes of the anomaly.  One troubling concern was that two
European experiments, the NUSEX\cite{NUSEX} and the
Frejus\cite{Frejus} Detectors, did not observe any anomaly.  Hence
some people suspected a peculiarity of water as a target or with the
employment of the Cherenkov radiation in vertex location.  However,
not only were the statistics of the European detectors relatively
small, but as indicated by more recent work from a similar type of
instrument in the US, the Soudan II detector\cite{SoudanII}, the
presence of a surrounding veto counter is vital for the more compact
type of slab detectors.  As well, the MACRO experiment\cite{MACRO} has
elucidated the non-negligible production of low energy (hundred MeV)
pions by nearby cascades in rock, which particles enter cracks in
non-hermetic detectors and appear to be neutrino interactions. In any
case the Soudan II with now significant exposure (4.6 kiloton-years)
finds an $R$ value close to that of SuperK (and IMB and
Kamioka)\cite{SoudanII}.

\begin{figure}
\includegraphics[width=.8\textwidth]{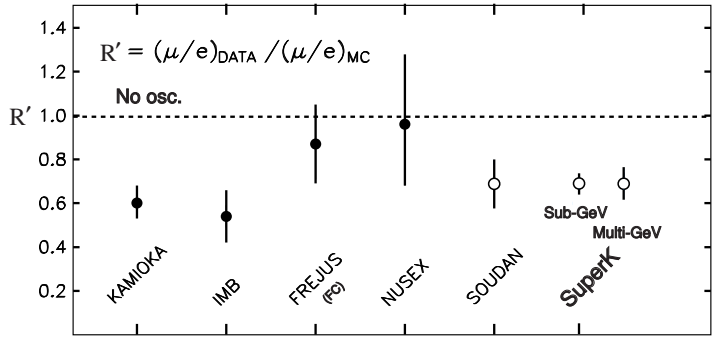}
\caption{The double ratio $R$ of muon to electron neutrino events,
data divided by expectations for various underground atmospheric
neutrino detectors. From A.~Mann\cite{Mann_99}}
\label{fig:rvalues}
\end{figure}

Figure \ref{fig:rvalues} shows these $R$ values for the several
experiments.  Note that under the assumption of no oscillations all
experiments should record $R = 1.0$, but if oscillations are taking
place, then the $R$ should be reduced but not necessarily the same
value for all experiments as it depends upon the energy range being
studied.  The European detectors did have higher thresholds which may
partly explain their failure to detect the anomaly.

\section{The Super-Kamiokande Revolution}

We now proceed to summarize the evidence for oscillations which while
depending largely upon the SuperK experiment, has important
confirmation and consistency of results from Soudan II and MACRO (and
consistency with previous smaller experiments as well, except as
discussed below).  Before going on it may be worthwhile to point out
what permitted the big break-through with SuperK, which may not be
obvious.  The increase in size of detector, from near kiloton fiducial
volumes for Kamioka and Soudan, and three kilotons for IMB to the
twenty-two kilotons of SuperK is not the whole story.  As will be seen
below, the most striking progress comes from the recording of muon
events with good statistics in the energy region above $1~GeV$. This
is due to detector linear dimensions as well as gross target volume:
muon events with energy more than $1~GeV$ and thus $5~m$ range were
not likely to be fully contained in the Kamioka detector (or the IMB
detector). SuperK in contrast has decent muon statistics up to almost
$5~GeV$, and this is turns out to be crucial.

The most important data to be discussed below is the ``fully
contained'' ($FC$) single-ring event sample, consisting of those
events in which both the neutrino interaction vertex and resulting
particle tracks remain entirely within the fiducial volume. For these
events the relativistic charged particle energy and direction are well
determined.  We shall use the notation $FC$ for the single ring
events, which are about 2/3 of the total, arising mostly from
quasi-elastic charged-current interactions in which the recoil nucleon
is not seen.  The multi-ring events have not yet been much used in
analysis due to the ambiguous interpretation of overlapping rings from
track segments in a Cherenkov detector (except as discussed below in
the case of tests distinguishing the muon neutrino's oscillating
partner). Moreover, the multi-pion final states are not modeled
reliably in simulations as yet, and there are further complications of
final state nuclear scattering as well.

There are also ``partially contained'' ($PC$) events, in which a muon
exits the fiducial volume from a contained vertex location.  Such
events are useful even though the total energy is not known, the
energy observed being a lower limit.  Of course this is the case even
with $FC$ events, though to a lessor degree, because the observed
particles are not of the same energy (or direction) as the incident
neutrino, which of course is what one would like to know.

\begin{figure}
\includegraphics[width=.8\textwidth]{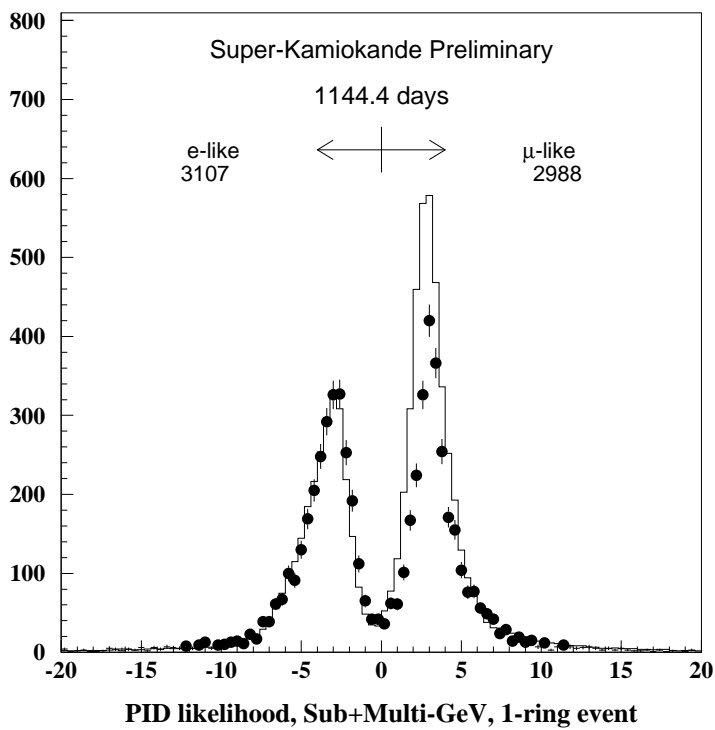}
\caption{Particle identification parameter distribution of the SuperK
fully contained single ring data, and Monte Carlo simulation, illustrating
the electron-like and muon-like separation\cite{Sobel_00}.}
\label{fig:pid}
\end{figure}

The particle types are identified by pattern recognition software, now
well tested and verified by experiment with known particle beams at
the accelerator\cite{Kasuga_96}.  Fortunately most of the contained
events show single (Cherenkov radiating) tracks in which the
identification is quite clean (at the 98\% level), as illustrated in
Figure \ref{fig:pid}.  To be clear and cautious we usually refer to
the reconstructed events as ``muon-like'' and ``electron-like'',
though a safe approximation is that these represent muon and electron
neutrino charged-current interactions.

\begin{figure}
\includegraphics[width=.8\textwidth]{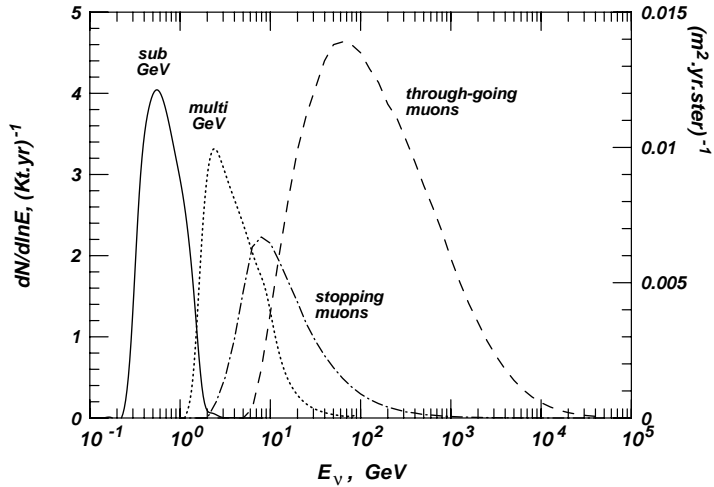}
\caption{Event rates as a function of neutrino energy for fully contained
events ($E <1.3~GeV$ and $E > 1.3~GeV$), stopping muons (and
similarly partially contained events), and through-going muons. From
Engel\cite{Engel_99}.}
\label{fig:nuenergy}
\end{figure}

The other two categories of events which we shall discuss are the
through-going upwards-moving muons ($UM$), produced by neutrino
interactions in the rock or outer detector, and which are coming from
directions below the horizon (as those from above the horizon can be
confused with down-going muons from cosmic ray interactions in the
atmosphere near overhead).  Another category of event is the
entering-stopping muon ($SM$).  It is useful that these event
categories probe approximately three different energy ranges of
neutrinos: $FC$ $\simeq 1~GeV$; $PC$ and $SM$ $\simeq 10~GeV$; $UM$
$\simeq 100~GeV$.  This is illustrated in Figure
\ref{fig:nuenergy}. It should be understood that as far as we know,
these neutrinos are all produced in the upper atmosphere by cosmic
ray interactions, and are reasonably well described by models in
content, energy, and angular dependence (to a few
percent)\cite{Stanev_99}.

We shall not take up limited space here with the description of the
SuperK detector, which is well documented elsewhere.  The interested
reader would do well to look at some of the theses from SuperK, which
are available on the web\cite{sk_pubs}.  The short summary is that the
SuperK detector consists of a large stainless steel cylinder ($37~m$
high by $34~m$ diameter inside the inner detector) containing a
structure holding 13,142 large (20 inch diameter) photo-multipliers .
With extremely high photo-cathode coverage (40\%), nearly an acre of
photocathode and ten times more pixels than any earlier instrument,
the instrument possesses a remarkable sensitivity of roughly eight
photoelectrons per MeV of deposited (Cherenkov radiating) energy.  The
latter permits detection of events down to $<5~MeV$, so for the
present discussion detection efficiency versus energy is not important
because the events we are discussing are all above $\simeq 100~MeV$.
The inner volume is also well protected by a $2$-$m$-thick,
fully-enclosing veto Cherenkov counter, populated by 1800 recycled IMB
(8 inch) photo-multipliers with wavelength shifting collars.  The
inner ``fiducial'' volume is further taken as $2~m$ inside the inner
photo-multiplier surface, resulting in the 22.5 kiloton volume used
for most reported data.

The SuperK oscillations claim was first formally presented to the
physics community in June 1998 at the $NEUTRINO98$ meeting in
Takayama, Japan.  The data were presented in several papers to the
community\cite{sk_sub,sk_multi,sk_ew}, building upon past data from
Kamioka\cite{jgl_kam} and IMB\cite{imb}, and culminating in the claim of
observation of oscillations of muon neutrinos, published in Physical
Review Letters in August 1998\cite{sk_nuosc}.  We now proceed to
review the evidence, which has changed little except for new
indications that the $\nu_{\mu}$ oscillating partner is probably the
$\nu_{\tau}$, and not a hypothetical sterile neutrino.

        \subsection{Up-Down Asymmetry}

\begin{figure}
\includegraphics[width=.8\textwidth]{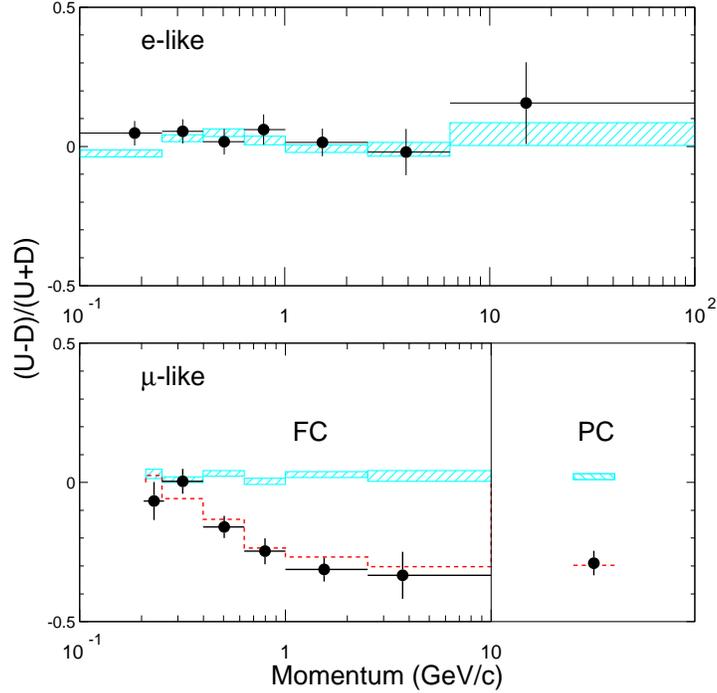}
\caption{The up-to-down asymmetry for muon (2486) and electron (2531) single
ring fully contained and partially contained (665) events in SuperK,
from 1144.4 days of live time (analyzed by 6/00), as a function of
observed charged particle momentum. The muon data include a point for
the partially contained events ($PC$) with more than about $1~GeV$.
The hatched region indicates no-oscillation expectations, and the
dashed line $\nu_\mu - \nu_\tau$ oscillations with $\Delta m^2 = 3.2
\times 10^{-3} eV^2$ and maximal mixing\cite{Sobel_00}.}
\label{fig:ud}
\end{figure}

One way to look at the $FC$ (and $PC$) data is in terms of a
dimensionless up-to-down ratio, difference over sum (which has
symmetrical errors in contrast to just up/down)\cite{updown}.
Downwards going neutrinos have flown $\sim 20-700~km$ while up going
neutrinos have traveled $\sim 700-10,000~km$. The angle between the
neutrino and observed charged lepton is on average of the order of
$40^o/\sqrt{E_{\nu}/GeV}$, and the typical observed energy is a half
the neutrino energy. Thus the mixing of the hemispheres of origin of
the events is important only for the lowest energies (below roughly
$400~MeV$).  This asymmetry quantity is exhibited as a function of
charged particle momentum in Figure \ref{fig:ud}, for both electrons
and muons, with the $PC$ data shown as well (for which we know only a
minimum momentum), from an exposure of 70.4 kiloton years in
SuperK. One sees that the electron data fit satisfactorily to no
asymmetry, whilst the muon data show strong momentum dependence,
starting from no asymmetry and dropping to about -1/3 above $1.3~GeV$.

From this Figure alone, without need for complex and often opaque
Monte Carlo simulations, assuming the cause of deviation from
uniformity to be due to neutrino oscillations, one can deduce that:

\begin{enumerate}

\item{ The source of the atmospheric neutrino anomaly is largely due
to disappearing muons, not excess electrons.}

\item{ There is little or no coupling of the muon neutrino to the electron
neutrino in this energy/distance range.}

\item{ The oscillations of the muon neutrinos must be nearly maximal
for the asymmetry to approach one third.}

\item{ The scale of oscillations must be of the order of
$1~GeV/200~km$, plus or minus a factor of several.}

\end{enumerate}

In fact, as seen by the dashed lines overlying the data points, the
simulations do produce an excellent fit to the muon neutrino
oscillation hypothesis, while the no-oscillations hypothesis is
strongly rejected.  The latter is so strong that statistical
fluctuations as the cause of the deviation are completely improbable;
one must look for systematic problems in order to escape the
oscillations explanation.

One concern for some people has been the fact that the asymmetry is
indeed maximal, which makes it appear that we are very lucky that the
earth size and cosmic ray energies are ``just so'' to produce this
dramatic effect.  This appears to this author to fall in the category
of lucky coincidences, such as the angular diameter of the moon and
sun being the same as seen from earth. There is another oscillations
related peculiar coincidence that the matter oscillation scale turns
out to be close to one earth diameter, and this depends upon the Fermi
constant and the electron column density of the earth.  The phase
space for ``coincidences'' is very large, and we humans are great
recognizers of such patterns.

        \subsection{Neutrino Flux Dependence Upon Terrestrial
        Magnetic Field}

The effect of the earth's magnetic field on the atmospheric neutrino
flux is a little complicated, but only important for very low
energies.  For example for energies of a few GeV, the magnetic field
provides some shielding from straight downwards going charged cosmic
rays in regions near the magnetic equator. For higher energies and
incoming trajectories near the horizon, the magnetic field still
prevents some arrival paths. As the SuperK detector location is not on
the magnetic equator the effect is not up-down symmetric, and this
spoils the symmetry otherwise expected from the neutrinos about the
horizontal plane.  However, the effects are mostly limited to neutrino
energies below about $1~GeV$, corresponding to cosmic ray primaries
below about $10~GeV$.  The picture is made a bit more complicated by
the earth's magnetic field not being a nice symmetrical dipole.
Fortunately there are good models of the magnetic field, and the
people who have made flux calculations take these effects into
account, though (in the past) largely through a simple cutoff momentum
versus location.  More recent calculations trace particles backwards
in the magnetic field and determine trajectories that escape to
infinity\cite{Stanev_98}.  Lipari has, however, recently shown that
the double humped cosmic ray spectra seen in the AMS experiment, with
a space borne magnetic spectrometer in low earth orbit, may be due to
particles in trapped orbits\cite{Lipari_00}.  Moreover, Lipari points
out that there are hints in the AMS data of a North-South asymmetry,
which could bias the neutrino flux calculations, and even pull the
derived value of $\Delta m^2$.  However, it should be emphasized that
the effect of such variation from simple expectations will only bias
the lowest energy data in SuperK (roughly below $400~MeV$), and
analysis has demonstrated that the results quoted herein are stable
against raising the acceptance energy for the data sample.

The SuperK group has published a paper\cite{sk_ew} examining the
azimuthal variation of the SuperK data ($\pm 30 \deg$ about the
horizon) for intermediate to higher energies ($400-3000~MeV$), in an
energy region where the calculations are thought to be
reliable. Indeed the SuperK data do exhibit significant variation from
uniformity while fitting the flux predictions very well, giving one
confidence in the modeling\cite{sk_ew}.

        \subsection{Natural Parameters for Oscillations: $L/E$}

\begin{figure}
\includegraphics[width=0.8\textwidth]{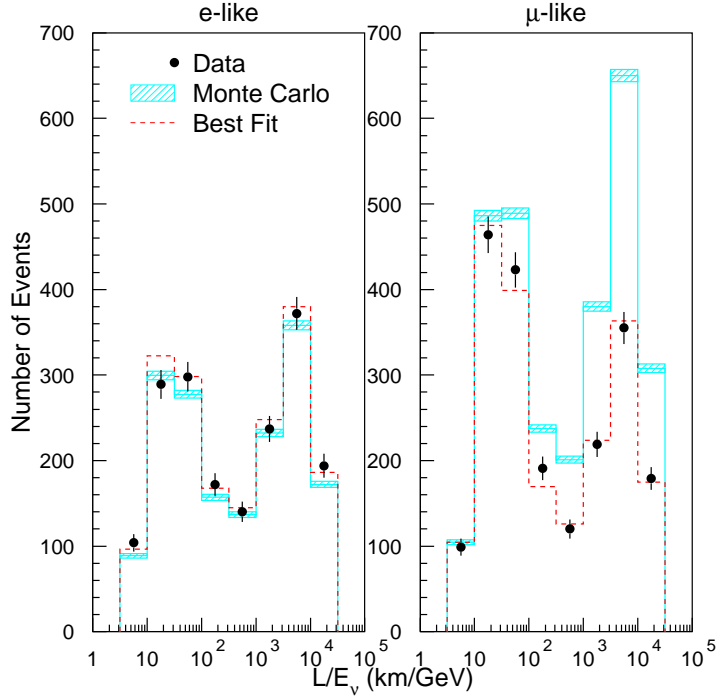}
\caption{The numbers of SuperK events observed compared to
predicted as a function of the natural oscillations parameter, $L/E$,
distance divided by energy.  The results are not normalized.  The two
peaks correspond to generally down-coming (left) and up-coming (right).
One sees that the muon deficit begins even in the upper hemisphere.
The shaded area indicates no-oscillations expectations, and the dashed
line in the fit for $\nu_\mu - \nu_\tau$ oscillations with maximal
mixing and $\Delta m^2 = 0.0032~eV^2$.\cite{Sobel_00}}
\label{fig:nloe}
\end{figure}

\begin{figure}
\includegraphics[width=0.8\textwidth]{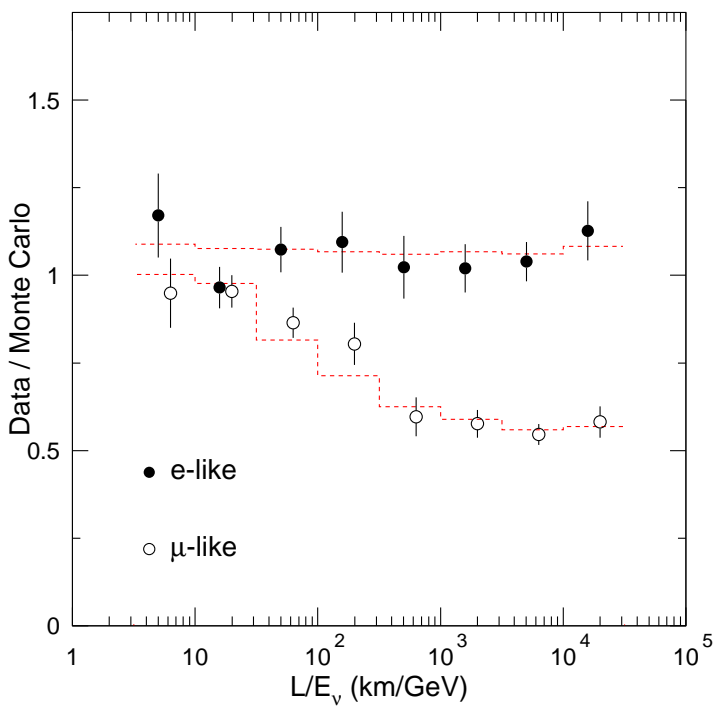}
\caption{The ratio of numbers of events observed compared to
predicted as a function of the natural oscillations parameter,
distance divided by energy.  The results are not normalized and
overall there is a slight excess (about 8\% compared to a systematic
uncertainty of 25\%) compared to expectations.  Electrons show no
evidence for oscillations, while muons exhibit a strong drop with
$L/E$.  This is consistent with $\nu_\mu - \nu_\tau$ oscillations with
maximal mixing and $\Delta m^2 = 0.0032~eV^2$, as indicated by the
dashed lines from the simulation.\cite{Sobel_00}}
\label{fig:loe}
\end{figure}

In an ideal world, one would assuredly study these data as a function
of distance divided by energy, $L/E$, since that is the parameter in
which one expects to see oscillatory behavior.  For two-neutrino
mixing with mass squared difference, $\Delta m^2$, and mixing angle
$\theta$, the probability of a muon neutrino of energy $E_\nu$
remaining a muon neutrino at distance $L$ is given by\cite{MNS}
\begin{equation}
P_{\mu \mu} = 1 - \sin^2(2\theta) \sin^2\left(1.27 \frac{\Delta m^2}{eV^2}
\frac{L}{km} \frac{GeV}{E_\nu}\right) .
\end{equation}
However, since we observe only the secondary charged particle's
energy and direction, badly smeared at the energies available
($L/E_\nu$ smeared by about a factor of two), plots in which one
would wish for visible oscillations can at best show a smooth slide
from the no-oscillations region to the oscillating regime.  This is
illustrated in Figures \ref{fig:nloe} and \ref{fig:loe}, where the
numbers of events, and the ratios of those numbers of events
observed to those expected with no-oscillations, are plotted versus
``$L/E$''\cite{Messier_99}, for muon and electron (type) events.
The updated data are preliminary from the SuperK 1144 day sample.

The plot is not ``normalized'', and we see somewhat of an excess of
electron type events overall ($+8\%$).  This is a little worrisome,
but acceptable since (as already noted) the absolute flux is uncertain
to a larger value.  In contrast to the electron data, the muon points
fall relative to no-oscillations expectations with increasing $L/E$
beyond about $50~km/GeV$, reaching a plateau at about one half their
initial value, consistent with maximal mixing. Muon (to tau) neutrino
oscillations in the Monte Carlo simulation are indicated by a dotted
lines, and fit the data reasonably well.

As noted, these data do not (and could not) show oscillations, due to
convolutions washing out the oscillatory behavior.  It was this smooth
fall, however, that caused the author and some colleagues to wonder if
another model might fit the data, one in which one component of the
muon neutrino decays rather than oscillates with distance.  Two
papers\cite{nudk1,nudk2} suggested neutrino decay to explain the
atmospheric neutrino anomaly.  I will not discuss details here, but
note that in order to construct a viable model we had to push on all
available limits, and invoke neutrino mass and mixing in any case.
Consequently such models do not pass the economy test of Occam's
Razor, though most annoyingly they remain not ruled out as yet.

One may note that detecting multiple oscillation peaks is not ruled
out in principle for detectors such as SuperK or Soudan II, employing
the atmospheric neutrinos.  It is a matter of recording the final
state of the muon neutrino charged current events, including nuclear
recoil, with sufficient accuracy and statistics.  Detectors such as a
liquid argon device of the ICARUS type are claimed to have the
resolution, if large enough.  Soudan II has apparently good enough
resolution to accumulate a ``golden sample'' in which the nuclear
recoil is detected, permitting reconstruction of incident neutrino
energy and direction.  Unfortunately Soudan II does not have enough
mass to achieve definitive statistics in a practical observing
period\cite{SoudanII}. Another possibility is that SuperK with enough
exposure and more highly developed analysis would be able to
accumulate an adequate sample of events in which the recoil proton is
detected above Cherenkov threshold.  At the moment none of the above
promise success.

Considering future experiments, the attempt to discern oscillations as
a function of $L/E$ is one area in which improvement may indeed be
made.  The MINOS\cite{MINOS} detector in Minnesota with a neutrino
beam from Fermilab, and the large detectors to be constructed in Gran
Sasso, ICANOE\cite{ICANOE} and OPERA\cite{OPERA}, detecting a neutrino
beam from CERN, give some hope of being able to yield oscillatory
plots.  A hypothetical detector, such as a megaton version of the
Aqua-RICH instrument studied by Ypsilantis and colleagues could have
the resolution to see a multi-peaked $L/E$
plot\cite{Ypsilantis,Learned_98}.  Nearer to technical development,
the proposed MONOLITH experiment would consist of a 30 kiloton pile of
magnetized iron and tracking detector layers, and employ the cosmic
rays to detect the first dip in $L/E$ in the upcoming muon flux
through the earth\cite{MONOLITH}.

        \subsection{Energy and Angle Variation}

The SuperK Collaboration's preferred method of fitting the ensemble
(single ring) $FC$ and $PC$ data is to employ a $\chi^2$ test to
numbers of events binned by particle type, angle, and energy, a total
of 70 bins.  The bin choices may seem a bit peculiar, but they have
historical precedent (they are as employed for Kamiokande) and though
not optimal for the new data set, this choice permits avoidance of any
statistical (or confidence) penalty for choosing arbitrary bins.  The
fit employs a set of parameters to account for potential systematic
biases.  Details cannot be presented here, but it has been shown that
the numerical results are quite insensitive to the selection of the
parameters or their supposed ``errors'' (except for the overall
normalization)\cite{pull}.  This method of systematic error handling
has been shown to be equivalent to employment of the correlation
matrix of parameters\cite{Messier_99}.

\begin{figure}
\includegraphics[width=.8\textwidth]{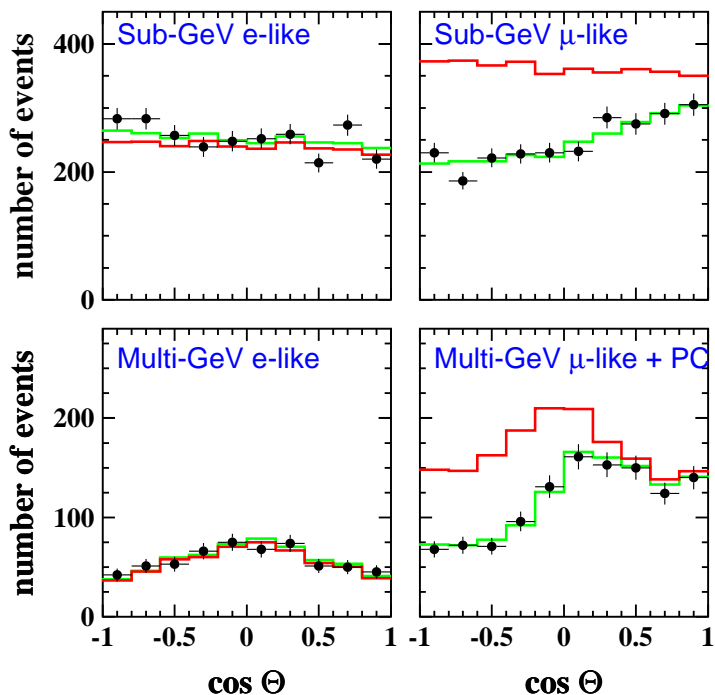}
\caption{Cosine of zenith angle distributions of the contained and
partially contained event data for two different energy ranges (above
and below $1.3~GeV$), electron and muon single ring events. 1144 live
days of SuperK data (preliminary analysis) are indicated by dots with
statistical error bars.  The solid line shows no-oscillations
simulation result, and the hatched line that for oscillations between
muon and tau neutrinos with best fit $\Delta m^2 = 0.0032~eV^2$ and
maximal mixing\cite{Sobel_00}.}
\label{fig:ae}
\end{figure}

Figure \ref{fig:ae} illustrates the data plotted for two energy
intervals ($sub-GeV$ and $multi-GeV$, less or more than $1.3~GeV$) for
single track events identified as either electron-like or muon-like.
The partially contained data is displayed with the multi-GeV muon
data.  The data are shown as a function of the cosine of the zenith
angle, with $+1$ being down-going.  One sees that the data very well
fit the curves from the Monte Carlo simulation, at the values gotten
from the grand ensemble fit, $\Delta m^2 = 0.003~eV^2$ and $sin^2( 2
\theta )=1.0$. The limit on $\Delta m^2$ is $0.002$ to $0.007~eV^2$,
and $\sin^2(2 \theta) >0.85$ at 90\% confidence level.

\begin{figure}
\includegraphics[width=.8\textwidth]{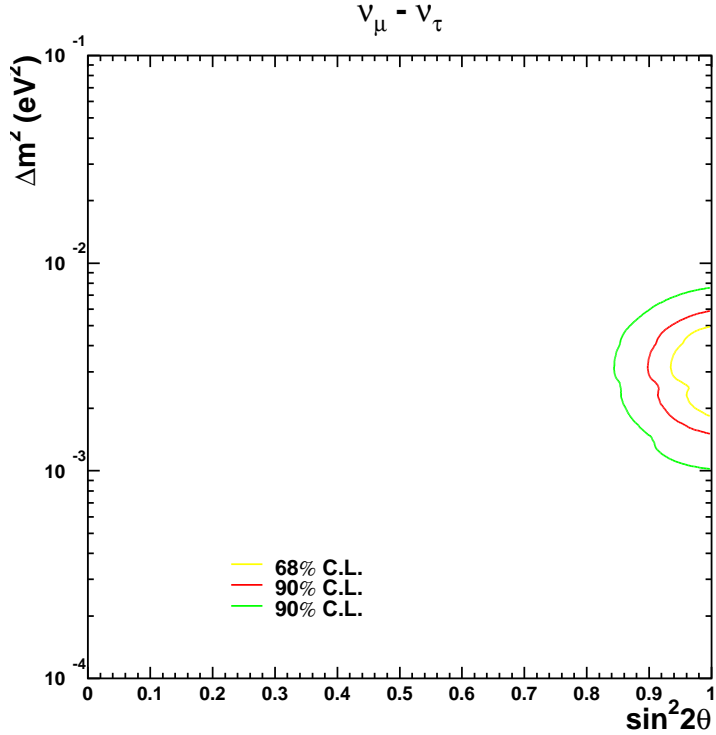}
\caption{Inclusion plot, showing the regions for 3 levels of
statistical acceptability in the plane of mixing angle and mass
squared difference for $\nu_\mu - \nu_\tau$ oscillations.  This is
from SuperK contained and partially contained event 1144 day data
preliminary analysis of June 2000\cite{Sobel_00}.}
\label{fig:excl}
\end{figure}

The results of the fits are often presented in terms of an inclusion
plot, showing an acceptable region(s) in the space of mixing angle
($\sin^2 2 \theta$) and mass squared difference ($\Delta m^2$), as
presented in Figure \ref{fig:excl}.  The $\Delta m^2$ value at
$\chi^2$ minimum has moved a little upwards with accumulated
statistics, though not much, (good news for long baseline experiments
anyway) but remains uncertain to about a factor of two.

It is noteworthy that the earlier indications of and constraints upon
oscillations parameters from Kamiokande, IMB and Soudan gave somewhat
larger values of $\Delta m^2$.  All of these results depended upon
fitting the $R$ value, since no angular distribution was discerned
(due to limited statistics and lower mean energy due to containment).
Later Kamiokande data did show angular variation in the PC data, but
not statistically compellingly. For some reason not fully understood,
the fits using $R$ alone all seem to yield higher values of $\Delta
m^2$.  If one has some deficit in muons without angular determination,
then one can fit that suppression with any $\Delta m^2$ above some
threshold value by choosing an appropriate mixing angle.  Thus the $R$
constraints are open ended upwards in $\Delta m^2$.  Perhaps there is
a systematic problem here due to predicted neutrino spectra, or
perhaps there is some physics yet to be elucidated.  This is not to
suggest that it seems possible for the preferred two-neutrino solution
to move much, but that more complex small effects at the $<10$\% level
could be superposed on the present simple solution.  Accelerator based
experiments should clarify this issue.

        \subsection{Muon Decay Events}

It is not often emphasized, but the original indication of the
anomaly, a deficit in stopped-muon decays ($\simeq 2.2~ \mu sec$ after
the initial neutrino event), remains with us, and constitutes a nice
alternate sample, almost independent and with quite different
systematics.  It is not so clean a sample (there are muon decays from
pions produced in electron CC and all flavor NC events) and the
statistics are lower, but the complete consistency of the muon decay
fraction remains a reassuring complement to the energy and angle
analysis employing track identification.

        \subsection{Through-Going and Entering-Stopping Muons}

Another cross check comes from the $UM$ and $SM$ samples, which are
particularly attractive because the source energies are factors of
$10$ and $100$ higher and the detector systematics rather different
(for example, the target is mostly rock not water).  A drawback to
these samples is that one is restricted to using muons arriving from
below the horizon, due to the overwhelming number of down-going cosmic
ray muons penetrating the mountain (at 50,000 times the rate in
SuperK).

In going from the earlier instruments to SuperK, however, the gain is
not so great (the $1200~m^2$ of SuperK being about a factor of three
more than the previously largest underground instrument, IMB, for
example), since the rate of collection of through-going muons depends
upon area not volume.  However, the much greater thickness of the
detector (and the efficient tagging of entering and exiting events in
the veto layer) yields many more entering-stopping ($SM$) events.

\begin{figure}
\includegraphics[width=.8\textwidth]{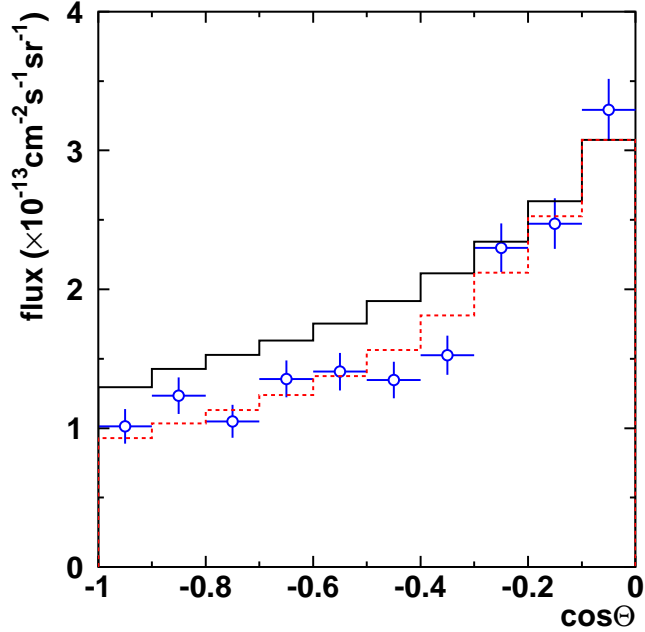}
\caption{The preliminary flux calculated from 1260 upcoming muons
during 1138 live days of SuperK data, with 9 events background
subtracted in bin nearest the horizon, $\cos(\Theta) = 0$.  Error bars
show statistical errors only. Expectations for no-oscillations (solid
line) and the best fit for $\nu_\mu - \nu_\tau$ oscillations with
$\Delta m^2 = 0.0032~eV^2$ and maximal mixing (dashed line), are
shown\cite{Sobel_00}. }
\label{fig:um}
\end{figure}

The flux angular distribution derived from 1260 $UM$ events from below
the horizon, each with more than $7~m$ track length in the detector,
is shown in Figure \ref{fig:um}, where one sees that the angular
distribution is nicely consistent with oscillations and not with
no-oscillations.  However, since much of the effect is close to the
horizon, where oscillations for the energies in question are just
setting in, one worries about contamination of the near horizon events
with in-scattered events from the much greater numbers of down going
muons.  There is no room for detail here, but SuperK does perform a
small background subtraction (9 events of 247, or 3.6\% in that one
bin) for events within 3 degrees of the horizon, but otherwise finds
no evidence for significant contamination\cite{sk_um}.

\begin{figure}
\includegraphics[width=0.6\textwidth]{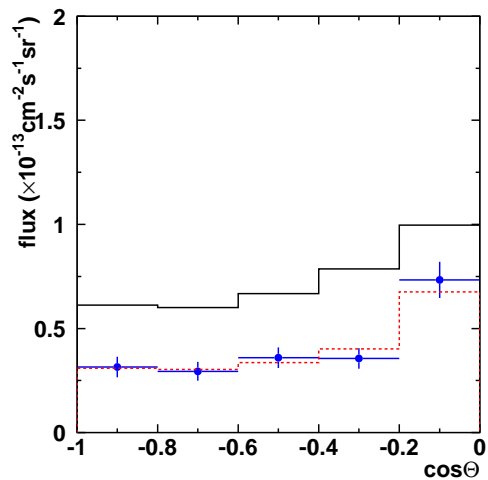}
\caption{The flux of stopping muons versus zenith angle in 1117 days
of SuperK data.  There are 311 events and a background of 21.4, with
track length greater than $7~m$, or about $1.6~GeV$.  The large
overall deficit in the data (dots with statistical error bars)
compared to the no-oscillations (solid line), is less significant than
it appears because of the 20\% uncertainty in absolute
flux\cite{Sobel_00}.}
\label{fig:sm}
\end{figure}

In SuperK the $SM$ sample was predicted to be 33-42\% of the $UM$
sample, as indicated in Figure \ref{fig:sm}, yet in fact SuperK sees
only about $24\% \pm 2\%$.  Fitting the data to the oscillation
hypothesis, one can make the now usual inclusion plot, which shows
that the $UM$ and $SM$ results are completely in accord with those
from the $FC$ and $PC$ data, see Figure \ref{fig:combi}.  However, as
the statistics are smaller and the physics leverage not as great, the
muon result does not add much to the $FC$ and $PC$ constraints, though
it does stiffen the lower bound in $\Delta m^2$. The joint fit to the
$UM$ and $SM$ data alone yields a $\chi^2/ndf = 35.4/15$ and $13/13$
for the cases of no-oscillations and $\nu_\mu - \nu_\tau$
oscillations.

One may note that earlier experiments, such as IMB, with a final
sample of 647 events, no veto counter and less mature flux
calculations, did not find any net deficit in the UM sample, nor
significant deviation from no-oscillations expectations.  A similar
case obtained with other smaller data sets.  Indeed, one may note that
on the strength of the SuperK UM data angular distribution alone, one
would hardly be making discovery claims.  All UM data from IMB and
Kamioka are and were in accord with the present results, but not
demanding of the oscillations conclusion.

There is a lengthy tale about an $SM/UM$ analysis from the IMB
experiment\cite{imb}, which claimed an exclusion region very close to
the now preferred solution.  The IMB stopping muon sample was small
and it was not clean due to lack of a veto layer.  More importantly
the interpretation seems to have been flawed due to older flux models
and Monte Carlo simulations.  Work is in progress to reassess the old
data with new flux calculations and an updated quark
model\cite{Casper_99}.  Thus there remains a cloud upon the horizon,
but one which may fade away upon reanalysis.  It might also be worth
recalling that in the 1980's the absolute rate of upcoming muons as
measured in the IMB, Kamioka and other detectors, agreed with
calculated rates employing then available flux calculations.  Also at
that time, peculiar angular distributions which fit no expectation
were reported at conferences from the MACRO and Baksan detectors.
These results all tended to give pause to claiming oscillations as the
resolution of the atmospheric neutrino anomaly.  These concerns were
swept away by the clean and statistically convincing $FC$ muon angular
distributions from SuperK.

        \subsection{The Muon Neutrino's Oscillation Partner}

\begin{figure}
\includegraphics[height=16cm]{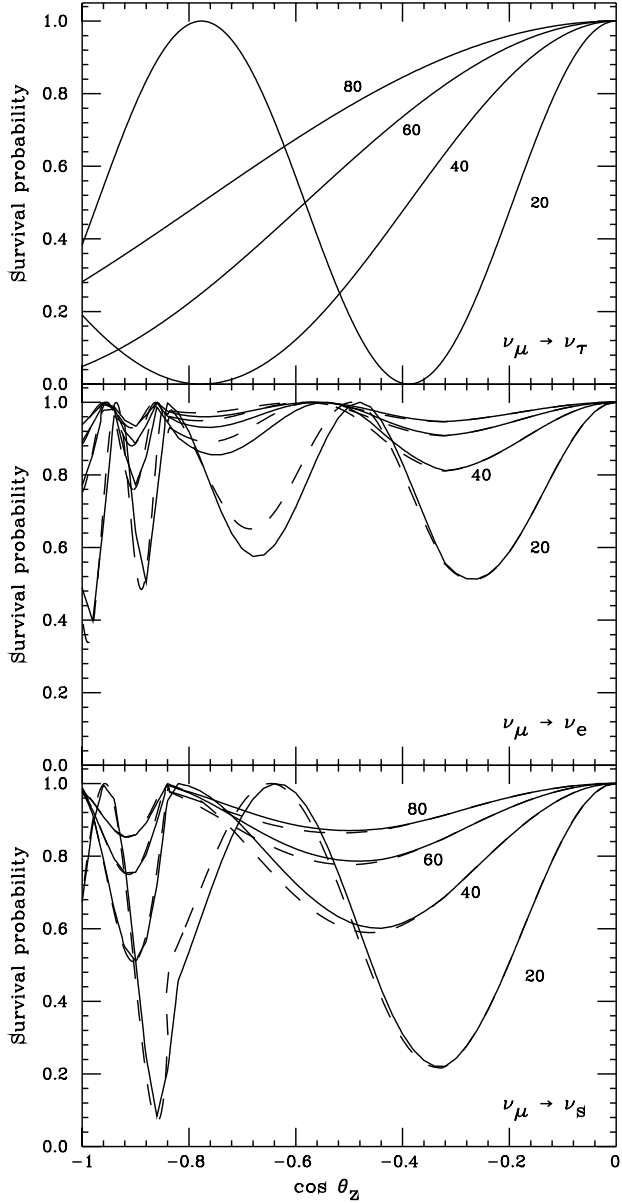}
\caption{Survival probability $P(\nu_\mu \to \nu_\mu)$ as a function
of the zenith angle in the cases of maximal mixing of $\nu_\mu$ with
$\nu_\tau$ (upper panel), $\nu_e$ (middle panel) and $\nu_s$ (lower
panel).  For $|\Delta m^2| = 5\cdot 10^{-3}$~eV$^2$ the curves
correspond to neutrino energies 20, 40, 60 and 80 GeV.  The dashed
curves are calculated with the approximation of constant average
densities in the mantle and in the core of the earth.  From
Lipari\cite{Lipari_98}.}
\label{fig:mattereffects}
\end{figure}

Given that the muon neutrino is oscillating, is it oscillating with a
tau neutrino or a new sterile neutrino which does not participate in
either the charged (CC) or neutral current (NC) weak interaction?
Fortunately there exist several means to explore this with SuperK
data.  The NC interactions should show an up-down asymmetry for
sterile neutrinos but not for tau neutrinos (since the NC interactions
for all ordinary neutrinos are the same).  Another avenue for
discrimination is that sterile neutrinos would have an additional
oscillation effect due to ``matter effects''.  The consequence would
be a unique signature in the angular distribution of intermediate
energy muons, as illustrated in Figure \ref{fig:mattereffects}.

Early SuperK efforts focused upon the attempt to collect a clean
sample of $\pi ^o$ events.  As it turns out, this was frustrated
because the rings (from the two decay $\gamma$s) cannot be separated
at energies above $\simeq 1~GeV$, and in net there are not so many
reconstructed events as to permit a good discrimination.  In fact the
absolute rate is consistent with expectations, but the cross section
is uncertain to about 20\% making the hint at tau coupling not
significant.  The $K2K$ experiment should soon measure this cross
section to perhaps 5\% however, making the $\pi^o$ rate a useful
discriminant.

More recently, tests have been devised employing a multi-ring sample
($MR$), the $PC$ event sample, and the $UM$ sample, all of which are
independent of the single ring $FC$ sample which yields the
strongest oscillation parameter bounds.

\begin{figure}
\includegraphics[width=1.0\textwidth]{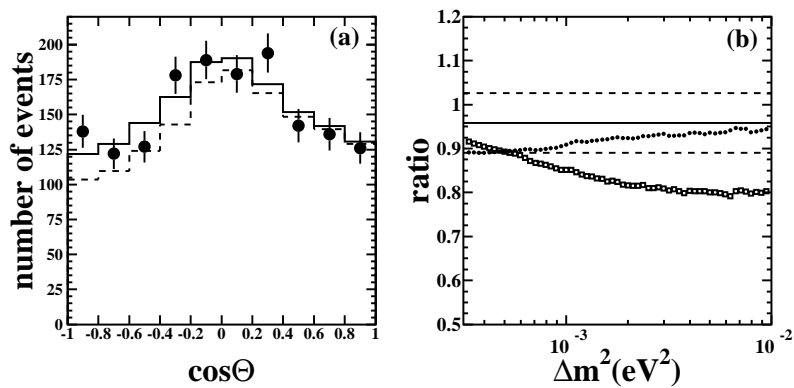}
\caption{(a) Zenith angle distributions of the multi-ring events
satisfying cuts as described in the text.  $cos\Theta$ is +1 for
down-going events.  The black dots indicate the data and statistical
errors. The solid line indicates the prediction for tau neutrinos, and
the dashed for sterile neutrinos with $(\Delta
m^2,sin^{2}2\theta)=(3.3\times 10^{-3} eV^{2},1.0)$.  These two
predictions are normalized by a common factor so that the number of
the observed events and the predicted number of events for $\nu_\mu
\leftrightarrow \nu_\tau$ are identical.  (b) Expected up/down ratio
as a function of $\Delta m^{2}$.  Horizontal lines indicates data
(solid) with statistical errors (dashed).  Black dots indicates the
prediction for tau neutrinos, and the empty squares for sterile
neutrinos, both for the case of maximal mixing\cite{tau-s}.}
\label{fig:mr}
\end{figure}

The $MR$ sample is cut by energy ($>1.5~GeV$) and the requirement of
the dominant ring being electron-like to enhance the NC content of the
sample.  A test parameter is constructed from the ratio of events from
within sixty degrees of the zenith and nadir.  This is illustrated in
Figure \ref{fig:mr}, where one sees consistency of $\nu_\tau$ and
disfavoring of $\nu_{sterile}$.

\begin{figure}
\includegraphics[width=1.0\textwidth]{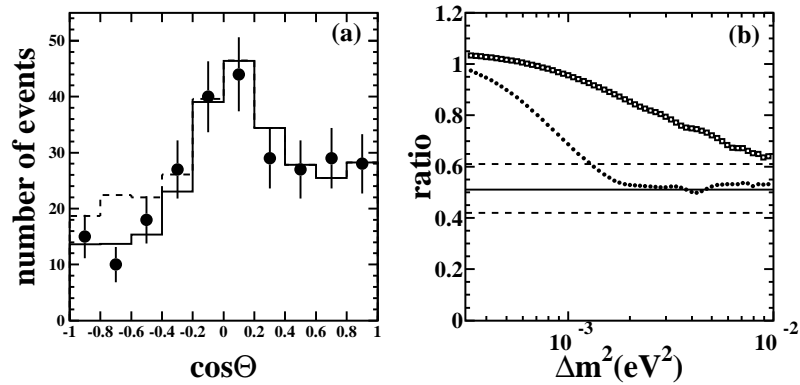}
\caption{(a) Zenith angle distributions of partially contained events
satisfying cuts of $E_{vis}>5~GeV$.  $\cos\Theta$ is +1 for down-going
events.  The black dots indicate the data and statistical errors. The
solid line indicates the prediction for tau neutrinos, and the dashed
for sterile neutrinos with $(\Delta m^2,sin^{2}2\theta)=(3.3\times
10^{-3} eV^{2}, 1.0)$ (b) Expected up/down ratio as a function of
$\Delta m^{2}$.  Horizontal lines indicate data (solid) with
statistical errors (dashed).  Black dots indicates the prediction for
tau neutrinos, and the empty squares for sterile neutrinos, both for
the case of maximal mixing\cite{tau-s}.}
\label{fig:pc}
\end{figure}

The $PC$ sample can be cut on energy (requiring $>4~GeV$) in order to
achieve a higher neutrino source energy, and the upwards going number
compared to downwards number of events. In this instance one is
seeking matter effects, and the results are shown in Figure
\ref{fig:pc}, indicating again a preference for $\nu_{\tau}$ over
$\nu_{sterile}$.

\begin{figure}
\includegraphics[width=1.0\textwidth]{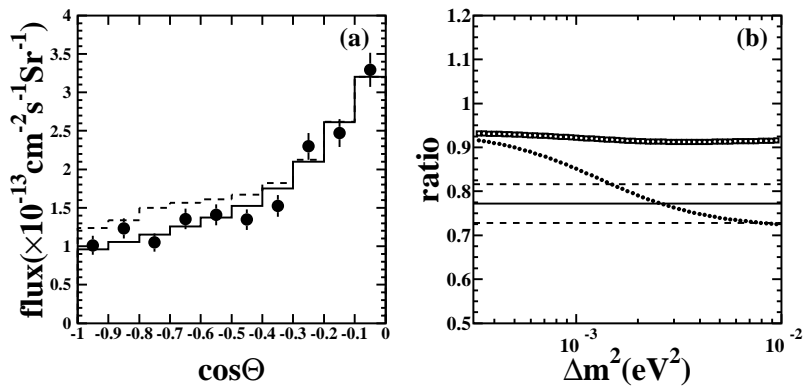}
\caption{(a) Zenith angle distributions of the upward moving
through-going muons. $cos\Theta$ is -1 for vertical up-going events.
The black dots indicate the data and statistical errors. The solid
line indicates the prediction for tau neutrinos, and the dashed for
sterile neutrinos with $(\Delta m^2,sin^{2}2\theta)=(3.3\times 10^{-3}
eV^{2},1)$ (b) Expected up/down ratio as a function of $\Delta m^{2}$.
Horizontal lines indicates data (solid) with statistical errors
(dashed).  Black dots indicates the prediction for tau neutrinos, and
the empty squares for sterile neutrinos, both for the case of maximal
mixing\cite{tau-s}.}
\label{fig:ums}
\end{figure}

For the muons, a near horizontal number can be compared to a number of
nearly straight upcoming events for another test of matter
oscillations.  This is presented in Figure \ref{fig:ums}.

\begin{figure}
\includegraphics[width=.6\textwidth]{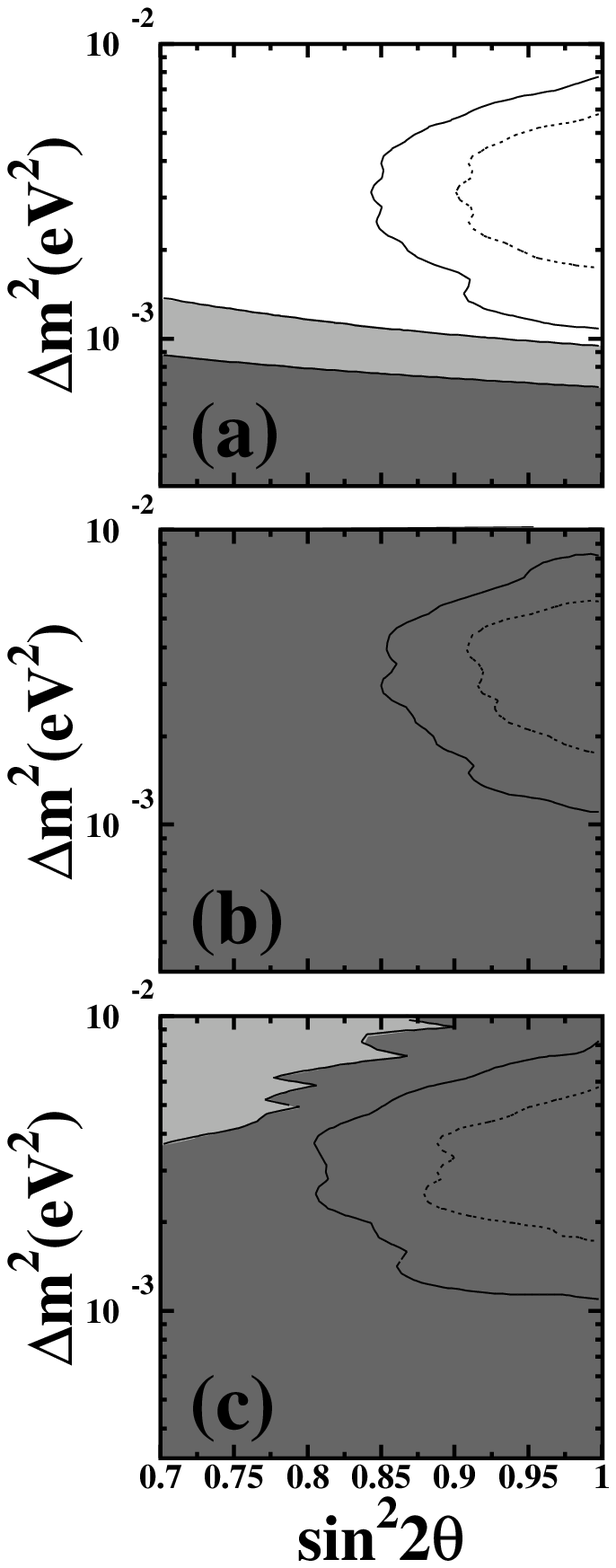}
\caption{Excluded regions for three alternative oscillation modes.
(a) $\nu_\mu \leftrightarrow \nu_\tau$, the light (dark) shaded region
is excluded at 90(99)\% C.L.,
(b)  $\nu_\mu \leftrightarrow \nu_{sterile}$ with $\Delta m^2 > 0$,
the whole region shown in this figure is excluded at the 99\% C.L.,
and (c) $\Delta m^2 < 0$, whole region is excluded at 90\% C.L., the
dark shaded region is excluded at more than the 99\% C.L..
The thin dotted (solid) line indicates the 90 (99)\% C.L.
allowed regions from the FC single-ring events\cite{tau-s}.}
\label{fig:taus}
\end{figure}

Finally the three tests are combined in a single $\chi^2$ test for the
case of $\nu_\mu \leftrightarrow \nu_\tau$ and the two cases for
$\nu_{sterile}$ heavier or lighter than $\nu_\mu$.  This is presented
in Figure \ref{fig:taus}, where one sees that the entire region in
mixing parameter space is eliminated for sterile neutrinos at more
than the 99\% confidence level, whilst the $\nu_\tau$ case fits
perfectly\cite{tau-s}

As to muon neutrinos coupling to electron neutrinos, SuperK can say
only that the $\Delta m^2$ is out of range on the low side or that the
sine of the mixing angle is less than about 0.1, if $\Delta m^2$ is
large.  As indicated in the earlier plots with up-down asymmetry,
there is surely not much mixing in this energy range.  One clever
scenario\cite{HPS} has the electron neutrino oscillation loss being just
compensated by muon neutrinos splitting their oscillations between
electrons and taus.  This seems to be ruled out by higher energy
SuperK data however\cite{Messier_99}.  Further discussions of (many)
other scenarios of oscillations can be found in the Chapter IX.

	\subsection{Sub-Dominant Oscillations}

For all of the foregoing we have considered only two flavor
oscillations.  A simple three flavor analysis can be made employing
the assumption that solar neutrino oscillations are driven by a much
smaller mass difference than found for $\nu_\mu - \nu_\tau$, $m_1 <
m_2 << m_3$.  Then, using the standard notation for three neutrino
MNS mixing matrix (see Chapter IX), the oscillation
probabilities can be written as:

\begin{equation}
P_{\nu_\mu \nu_\mu} = 1 -4\sin^2(2\theta_{23}) \cos^2(\theta_{13})
(1 -sin^2(\theta_{23})cos^2(\theta_{13})) \sin^2(1.27\Delta m^2 L/E),
\end{equation}

\begin{equation}
P_{\nu_\mu \nu_\tau} = \cos^4(\theta_{13}) \sin^2(2\theta_{23})
\sin^2(1.27\Delta m^2 L/E),
\end{equation}

\begin{equation}
P_{\nu_\mu \nu_e} = \sin^2(2\theta_{13}) \sin^2(\theta_{23})
\sin^2(1.27\Delta m^2 L/E),
\end{equation}

\begin{equation}
P_{\nu_e \nu_\tau} = \sin^2(2\theta_{13}) \cos^2(\theta_{23})
\sin^2(1.27\Delta m^2 L/E),
\end{equation}

\begin{equation}
P_{\nu_e \nu_e} = 1 - \sin^2(2\theta_{13}) \sin^2(1.27\Delta m^2 L/E),
\end{equation}

where $\Delta m{^2} = m{_3}^2 - m{_2}^2$, and we neglect any
consideration of $CP$ or $CPT$ violation.

\begin{figure}
\includegraphics[width=.6\textwidth]{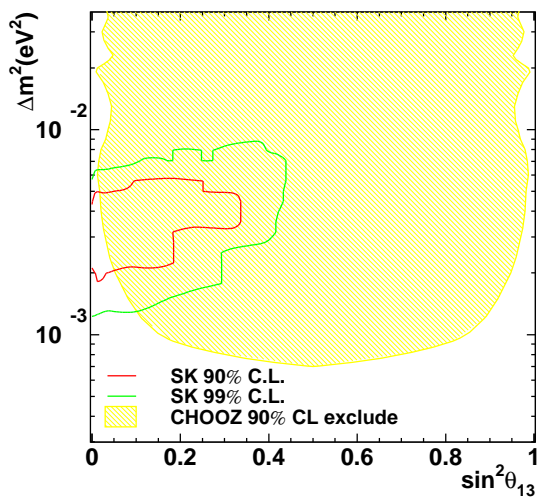}
\caption{Three neutrino fits to the SuperK data showing 90\% and
99\% allowed regions.  The shaded region is the 90\% exclusion
region from the CHOOZ experiment\cite{Obayashi_00}. See text for
qualifications.}
\label{fig:th13}
\end{figure}

The SuperK FC and PC data 990 day data has been fitted with these
equations\cite{Obayashi_00}, and the limits on $\sin^2(2\theta_{13})$
are less than 0.25 at 90\% C.L., with $\Delta m^2 = 0.003~eV^2$.
The best fit value lies at $sin^2(\theta_{13}) = 0.03$,
$sin^2(\theta_{23}) = 0.63$.  At this $\Delta m^2$ the CHOOZ results
allow either $sin^2(\theta_{13}) < 0.03$ or $> 0.97$, so the latter
is eliminated by the SuperK results, as illustrated in Figure
\ref{fig:th13}.

	\subsection{Non-Standard Oscillations}

\begin{figure}
\includegraphics[width=.8\textwidth]{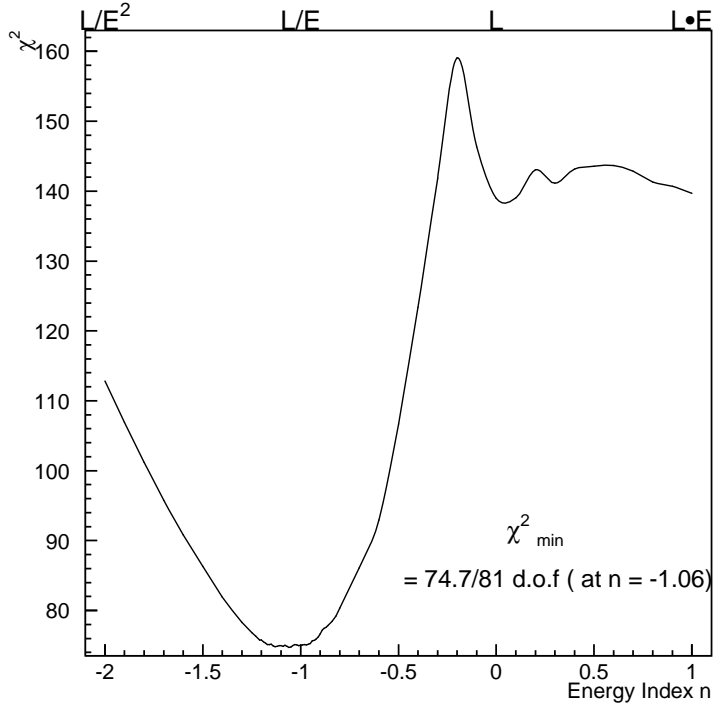}
\caption{The variation of $\chi^2$ with index $n$ in oscillating
phase for muon neutrinos, as in $const*L*E^n$. $n$ = -1 corresponds to
ordinary oscillations.  One sees that non-standard solutions with $n$
of 0 and 1 are strongly disfavored\cite{Sobel_00}.}
\label{fig:enn}
\end{figure}

As discussed in Chapter IX and elsewhere\cite{Pakvasa_99}, there are
models of neutrino oscillations which result from gravitational
splitting, violations of Lorentz invariance, and so on, which result
in oscillation with a phase proportional to, say, $L \times E$ instead
of $L/E$, or even with no energy dependence.  The SuperK group has
presented a fit to the data as a function of
\begin{equation}
 P_{\nu_\mu \nu_\tau} = \sin^2 (2\theta) \sin^2 (\beta L E^n)
\end{equation}
where $n$, $\beta$ and $\sin^2(2\theta)$ (0.7 to 1.3) are varied.
As indicated in Figure \ref{fig:enn}, the exotic solutions
are strongly disfavored, while the normal function is perfectly
acceptable, with $n = -1.06 \pm 0.14$.

        \subsection{Hypotheses to Explain Anomaly}

We conclude with a summary, presented in Table \ref{tab:hyp}, of all
hypotheses put forth to explain the atmospheric neutrino anomaly.
Space does not permit a full discussion here, but it is the case that
the SuperK data now have eliminated almost all alternate hypotheses to
explain the results.

The atmospheric neutrino flux calculations cannot be producing the
anomaly since we see the effect otherwise unexplained in the muon
zenith angle distribution.  The cross sections cannot produce such a
geographical effect, even if one could find some lepton universality
breaking phenomenon.  Particle identification has been heavily studied
and verified at the accelerator with a 1000-ton Cherenkov detector
tank\cite{Kasuga_96}.  Entering backgrounds, which would produce
effects clustering near the outer walls, show no evidence for
contribution to the anomaly in the SuperK data.  Detector asymmetrical
response is ruled out by the observed symmetry of the electron data,
as well as detailed calibration studies with isotropic sources.

Extra-terrestrial neutrinos cannot be the culprits (unfortunately),
since we see that it is a deficit of muon neutrinos causing the
anomaly, not an excess of electron neutrinos.  Proton decay is ruled
out similarly due to geographical dependence of the deficit, and the
extent of the anomaly to too high an energy (ruling out
neutron-anti-neutron oscillations as the source as well). The decay of
a muon neutrino component has been discussed, and while seemingly
unlikely is not totally ruled out as yet.  Anomalous absorption of
muon neutrinos by the earth, correlated with an exponential in the
column density through the earth, is ruled out as well.  This follows
because the anomaly is not dominated by that small part of the solid
angle going through the earth's core, whereas we see the muon deficit
starting even above the horizon.  Electron neutrino mixing is ruled out as
the dominant effect, again due to the muon angular distribution.
Sterile neutrinos do not fit the data as discussed above.  As
discussed in the previous section, non-standard oscillations are
also ruled out.

The only hypothesis which survives, and which fits all the evidence,
is that muon neutrinos maximally mix with tau neutrinos with a $\Delta
m^2$ in the range of $2-7 \times 10^{-3}~eV^2$.  It is noteworthy to
this author that in all the tests made on the data sample, there
appears to be great stability in the results against variations of all
the parameters explored.

\begin{table}
\caption{List of hypotheses invoked to possibly explain the
atmospheric neutrino anomaly. The first 3 columns are criteria
available prior to SuperK, and the last 4 after the 1998 SuperK
publication. The hypotheses divide into 5 systematics issues and 8
potential physics explanations. As indicated in the text, the only
remaining likely hypothesis is the oscillation between muon and tau
neutrinos. The ``$\times$'' schematically indicates which evidence
rules out the hypothesis in that row.}
\label{tab:hyp}
\vspace{2mm}
\begin{center}
\begin{tabular}{| l || c  |  c   |  c  ||  c   |   c  |   c  |  c   |}
\hline
 {\it Evidence}   &      &{\it Old} &  &      &     &{\it New}&    \\
\hline
                  & $R<1$ & $\mu~decay$ & $Vol$ & $R<1$ & $A_e$ & $A_{\mu}$
 & $R(L/E)$ \\
  Hypothesis      & $(E <1$ & $Frac$ & $Frac$ & $(E >1$ & $\simeq0$
 & $<0$ & $\simeq0.5$ \\
                  & $GeV)$ &      &       & $GeV)$ &      &       &       \\
\hline \hline
                  &       &       &       &       &       &       &       \\
  Atm. Flux Calc. &{\bf$\times$}&$\surd$&$\surd$&{\bf$\times$}&$\surd$&{\bf$\times$}&{\bf$\times$}\\
                  &       &       &       &       &       &       &       \\
  Cross Sections  &{\bf$\times$}&$\surd$&$\surd$&{\bf$\times$}&$\surd$&{\bf$\times$}&$\surd$\\
                  &       &       &       &       &       &       &       \\
  Particle Ident. &$\surd$&{\bf$\times$}&{\bf$\times$}&$\surd$&$\surd$&{\bf$\times$}&$\surd$\\
                  &       &       &       &       &       &       &       \\
  Entering Bkgrd. &$\surd$&$\surd$&{\bf$\times$}&$\surd$&$\surd$&{\bf$\times$}&$\surd$\\
                  &       &       &       &       &       &       &       \\
  Detector Asym.  &$\surd$&$\surd$&{\bf$\times$}&$\surd$&{\bf$\times$}&$\surd$&$\surd$\\
                  &       &       &       &       &       &       &       \\
\hline
                  &       &       &       &       &       &       &       \\
  X-Ter. $\nu_e$  &$\surd$&$\surd$&$\surd$&$\surd$&$\surd$&{\bf$\times$}&{\bf$\times$}\\
                  &       &       &       &       &       &       &       \\
  Proton Decay    &$\surd$&$\surd$&$\surd$&{\bf$\times$}&$\surd$&{\bf$\times$}&{\bf$\times$}\\
                  &       &       &       &       &       &       &       \\
$\nu_{\mu}$ Decay &$\surd$&$\surd$&$\surd$&$\surd$&$\surd$&$\surd$&$\surd$\\
                  &       &       &       &       &       &       &       \\
$\nu_{\mu}$ Abs.  &$\surd$&$\surd$&$\surd$&$\surd$&$\surd$&$\surd$&{\bf$\times$}\\
                  &       &       &       &       &       &       &       \\
$\nu_{\mu}-\nu_e$ &$\surd$&$\surd$&$\surd$&$\surd$&{\bf$\times$}&$\surd$&$\surd$\\
                  &       &       &       &       &       &       &       \\
$\nu_{\mu}-\nu_s$ &$\surd$&$\surd$&$\surd$&$\surd$&$\surd$&{\bf$\times$}&$\surd$\\
                  &       &       &       &       &       &       &       \\
Non-Stand. Osc.   &$\surd$&$\surd$&$\surd$&$\surd$&$\surd$&$\surd$&{\bf$\times$}\\
                  &       &       &       &       &       &       &       \\
$\nu_{\mu}-\nu_\tau$&$\surd$&$\surd$&$\surd$&$\surd$&$\surd$&$\surd$&$\surd$\\
                  &       &       &       &       &       &       &       \\
\hline
\end{tabular}
\end{center}
\end{table}

\subsection{Results from Soudan II}

\begin{figure}
\includegraphics[width=1.0\textwidth]{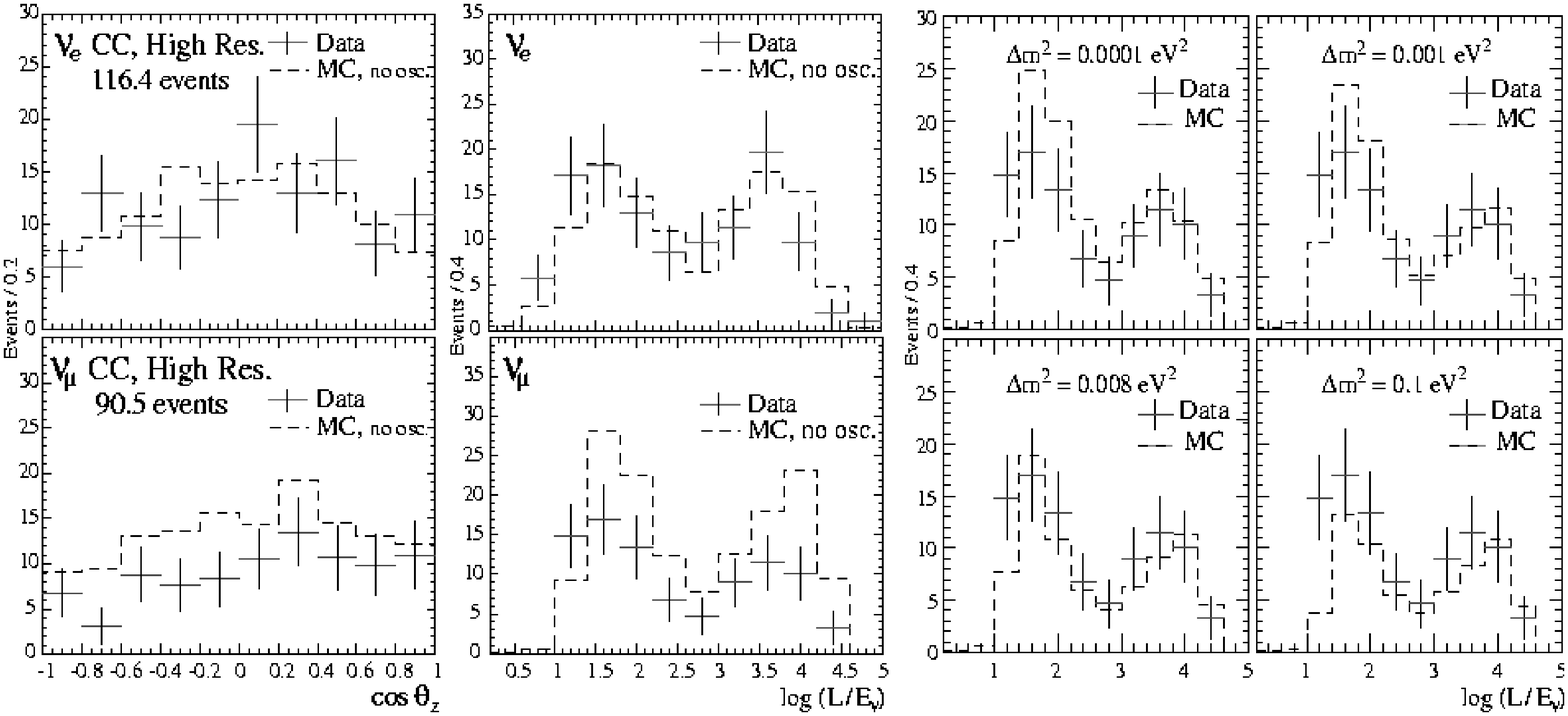}
\caption{ ({\it left}): Distributions in $\cos\theta_z$ for the $\nu_e$ and
$\nu_\mu$ flavor event samples. Data (crosses) are compared to the
null oscillation MC (dashed histogram) where the MC has been rate
normalized to $\nu_e$ data.  ({\it middle}): Distributions of $\log
(L/E_{\nu})$ for $\nu_e$ and $\nu_\mu$ charged current events compared
to the neutrino MC with no oscillations; the MC has been
rate-normalized to the $\nu_e$ data. ({\it right}): Comparison of
$L/E_{\nu}$ distribution for $\nu_\mu$ data (crosses) and expectations
from neutrino oscillations for four $\Delta m^2$ values, with
$\sin^2(2\theta) = 1.0$. From Mann\cite{Mann_99}}
\label{fig:soudan}
\end{figure}

The Soudan 2 detector located in an old iron mine in Minnesota, USA,
consists of a vertical slab, fine-grained tracking calorimeter of 963
tons total mass.  The cavern has a surrounding layer of proportional
tubes, 2 or 3 layers thick on all sides.  Data have been reported from
4.6 kiloton years exposure\cite{Mann_99}, as illustrated in Figure
\ref{fig:soudan}.  The contained events plotted are selected for
lepton energy $>700~MeV$ with no visible nuclear recoil, or visible
energy $>700~MeV$ and summed momentum $>450~MeV/c$ and lepton momentum
$>250~MeV/c$.  The energy resolution is of order 20\%, and the angular
resolution of order 20-30 degrees.  In the figure the predicted number
of events has been normalized to the electron total.  One can see that
with only of the order of 100 events of each type, the statistical
significance is not great, but the depletion of upcoming muon events
is evident.  The fits to oscillations parameters are included below,
in Figure
\ref{fig:combi}.

\subsection{Results from MACRO}

The MACRO detector, built primarily to seek monopoles, possesses a
significant capability, with effective mass of 5.3 kilotons, to detect
through-going and stopping muons as well as contained and partially
contained neutrino interactions\cite{Surdo_99}.  The instrument,
located in the deep underground Gran Sasso National Laboratory in
Italy, consists of horizontal planes of a tracking instrument.

\begin{figure}
\includegraphics[width=0.8\textwidth]{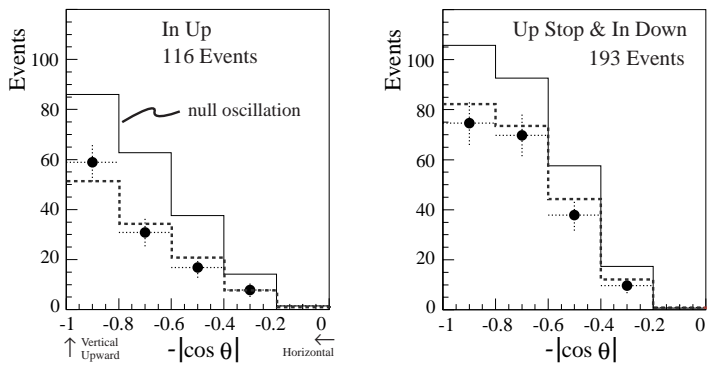}
\caption{Distributions in $\cos{\theta}$ for MACRO partially
contained events.  The data (solid circles) are seen to fall below the
null oscillation expectation in every bin of both samples. The dashed
line shows expectations for maximal mixing and $\Delta m^2 =
0.0025~eV^2$.  From Surdo\cite{Surdo_99}.}
\label{fig:macro_pc}
\end{figure}

Figure \ref{fig:macro_pc} shows the results of analysis of partially
contained data, for which up and down cannot be distinguished, but
which shows a clear deficit compared to expectations with
no-oscillations.  The acceptance of such a planar instrument is small
near the horizon so most of the effect is from the nearly vertical
events.  The In-Up sample has 116 events and exhibits a depletion of
$0.57 \pm 0.16$ compared to expectations.

\begin{figure}
\includegraphics[width=0.8\textwidth]{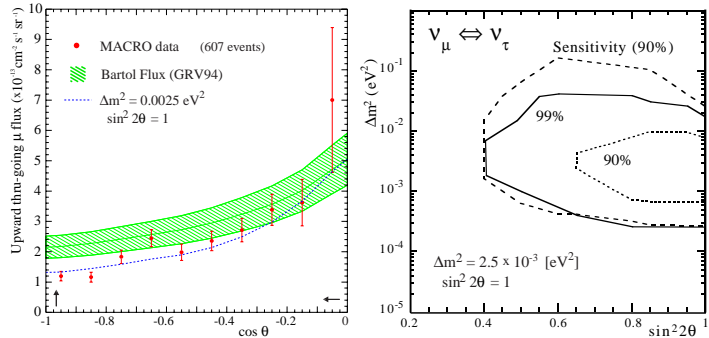}
\caption{ a) (left): The angular distribution of upward
through-going muons observed in MACRO. The data distribution (solid
circles) differs from the null oscillation expectation (shaded band)
in shape and in rate.  b) (right): The neutrino oscillation allowed
regions obtained by MACRO from the upward through-going
muons. Confidence level and experimental sensitivity boundaries are
calculated using the Feldman-Cousins method. From \cite{Surdo_99}.}
\label{fig:macro_um}
\end{figure}

The MACRO detector also has a significant sample of upwards
through-going muons, as illustrated in Figure \ref{fig:macro_um}.
Note that the muon energy threshold is low, in the hundred MeV range
(varying with angle and entry location).  The overall depletion
(data/expectations) is $0.74 \pm 0.03 \pm 0.04 \pm 0.12$, where the
last term reflects uncertainty in flux and cross
section\cite{Surdo_99}.  While the fit to oscillations appears not to
be perfect and the minimum in $\chi^2$ lies in the unphysical region,
the confidence level boundaries shown on the right in Figure
\ref{fig:macro_um} indicate consistency with the SuperK results.

\subsection{Combined Evidence}

\begin{figure}
\includegraphics[width=0.8\textwidth]{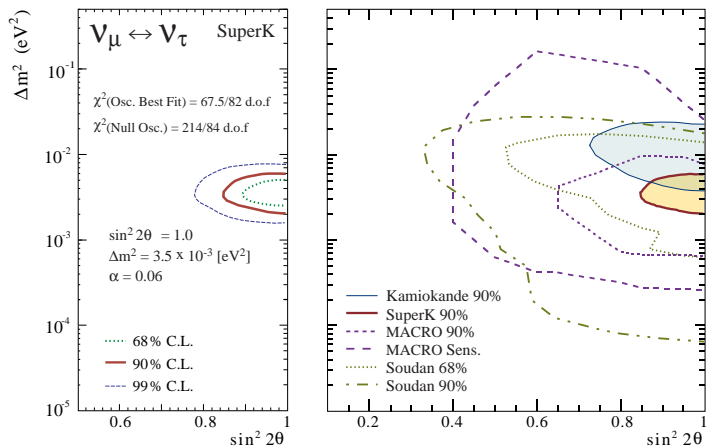}
\caption{a) (left): Allowed regions obtained by Super-K based upon
$\chi^2$ fitting to FC and PC single ring events, plus upward
stopping muons, plus upward through-going muons.  b) (right):
Oscillation parameter allowed regions from Kamiokande (thin-line
boundary), Super-K (thick-line boundary), MACRO (dashed boundaries),
and Soudan 2 (dotted and dot-dashed boundaries). From
Mann\cite{Mann_99}.}
\label{fig:combi}
\end{figure}

In Figure \ref{fig:combi} on the left is the combined fit of FC and PC
(848 live days) plus UM (923 live days) plus SM (902 live days) from
SuperK, with results as indicated, constraining the oscillation
parameters to be roughly $0.002 < \Delta m^2/eV^2 <0.006$ and $0.85 <
\sin^2 (2 \theta) < 1.0$, with the fit minimum in the physical
region at maximal mixing and $0.0035~eV^2$.  The boundaries for
MACRO and Soudan 2 are shown overlying the SuperK results in the
right hand panel.  One can see that the results of the largest
instruments now reporting atmospheric neutrino data, water Cherenkov
and two dissimilar tracking calorimeters, all agree on muon neutrino
disappearance, and are consistent with oscillation between muon and
tau neutrinos. I cannot do better than to quote Tony Mann from his
1999 Lepton-Photon conference plenary talk\cite{Mann_99}: ``I
propose to you that congratulations are in order for the researchers
of Kamiokande and of Super-K and more generally, for the
non-accelerator underground physics community. For
Fig.~\ref{fig:combi}b, Ladies and Gentlemen, is the portrait of a
Discovery - the discovery of neutrino oscillations with two-state
mixing.''

\subsection{Long Baseline Results}

The K2K experiment has been in operation for over one year at this
writing.  The $12~GeV$ KEK proton synchrotron has been equipped with a
neutrino line, and a double detector has been built at about $100~M$
range, on the KEK campus.  Neutrinos are aimed at the SuperK detector
at a distance of $250~km$, and events have been recorded\cite{K2K}.
Note that a $1~GeV$ neutrino would be near first minimum after 250 km
if the $\Delta m^2 = 0.005~eV^2$.

At the time of this writing preliminary results are available, and it
has been reported in conferences that the rate is low, consistent with
the SuperK results, and inconsistent with no-oscillations at the
$2~\sigma$ level\cite{Nakamura_00}.  From the period of June 1999
through March 2000, the group received about $1.7 \times 10^{19}$
protons on target (17\% of proposal request), during the equivalent of
about five months running.  The events were easily extracted from a
GPS synchronized $1.5~\mu sec$ time window relative to the $12~GeV$
KEK proton synchrotron fast beam spill ($2.2~sec$ repetition period
and about $5 \times 10^{12}$ protons on target per spill). During this
run 17 associated events were collected in the SuperK 22.5 kiloton
fiducial volume, when $29.2 \pm 3.1$ were expected for no
oscillations. The expected numbers are 19.3, 12.9 and 10.9 for 3, 5,
and 7 $\times 10^{-3}~eV^2$.  Of these SuperK events, 10 are single
ring, one of which is identified as electron-like (about equal to
expectations).

Given the low energy of the neutrino beam ($\sim 1~GeV$), and
frustratingly low data rate at SuperK probably not much more can be
expected from this experiment than a simple confirmation of the SuperK
atmospheric neutrino results.  The full proposal-run would collect 174
or so events with no oscillations, and assuming oscillations, about
half that.  Of those, about half should be single ring muons, leaving
perhaps 40-50 events from which to deduce the arriving neutrino
spectrum. Given that the accelerator delivers beam about half a year
each calendar year, one can see that it will take several years to
achieve definitive results.

\section{Implications}

The ramifications of the explication of the atmospheric neutrino
anomaly in terms of neutrino mixing and thus neutrino mass, are great
and span the known realms of fundamental physics from large to small.
We have not discussed in this chapter the links to solar
neutrinos\cite{Haxton}, nor the LSND results\cite{Caldwell}.
Certainly there is no conflict between the atmospheric muon neutrino
results and the possible (nay likely) solar oscillations.  If,
however, both solar and the LSND results are correct, then we have
surely some interesting physics to untangle, as it is generally
admitted that no simple three neutrino model can incorporate all three
neutrino anomalies, and that new degrees of freedom would be required
(see Chapter IX).  From the evidence presented in this chapter alone,
however, we can make some far reaching, perhaps paradigm shifting,
conclusions.

        \subsection{Astrophysics and Cosmology}

The implications of the oscillations results have been explored in
other chapters in this book, so here we only outline those relative to
the muon neutrino oscillations.  First, it appears that neutrinos with
summed masses of the order of $0.1~eV$ will not make any major
contribution to resolving the dark matter quandary.  Nonetheless with
a ratio of 2 billion to one for photons (and neutrinos) to nucleons
from the Big Bang, even such a small neutrino mass may be greater in
total than all the visible stars in the sky.  Hence while one must
account for neutrino mass in further cosmological modeling, neutrinos
are not likely to constitute the bulk of the ``missing matter''.
However if the neutrinos should be nearly degenerate in mass and all
have masses in the range near $1~eV$ (and hence we are observing only
small splittings with the oscillations), then neutrino mass may
dominate the universe.  While neutrinos are not favored by
astrophysical modelers (fitting the spatial fluctuations in the cosmic
microwave background for example), large neutrino masses are not ruled
out ($\Sigma M_i \leq 6~eV$, $H_o = 65~km/s/Mpc$, $i =
1,2,3$)\cite{Primak}.  Nearly degenerate neutrino masses would not
present a consistent picture with the quark and charged lepton masses,
which make large mass jumps between generations.  But who knows?  We
do not have a viable GUT with mass predictions\cite{Mohapatra}, so an
open mind is appropriate.

The other major area of significance, perhaps of the deepest
significance, has to do with baryogenesis, the origin of the
predominance of matter by one part per billion over anti-matter at Big
Bang time.  There are claims that the old idea of accumulation of net
baryon number via CP violation in the quark sector, while satisfying
the Sakharov conditions\cite{Sakharov_67}, may not suffice from the
early stages of universe expansion\cite{Kuzmin_70}. If that is indeed
the case, it may be that neutrinos provide the avenue for net baryon
asymmetry generation, with the expression into hadrons becoming
manifest relatively late in the game at electroweak phase
transition\cite{Akhmedov_98}. CP and even CPT violation in the
neutrino sector, as yet almost unconstrained, could have dramatic
implications.

Neutrino masses and possible sterile neutrinos have also been
invoked to help resolve problems in understanding heavy element
synthesis in supernovae\cite{Fuller}.

        \subsection{Theoretical Situation: Why So Important}

\begin{figure}
\includegraphics[width=.8\textwidth]{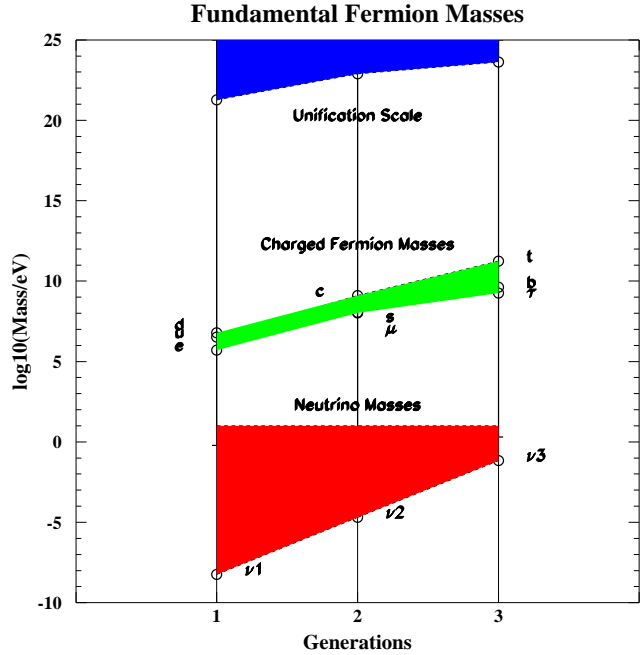}
\caption{The masses of the fundamental fermions. The lower shaded region
indicates the rough range allowed by present oscillations results,
with the upper boundary for nearly degenerate neutrino masses and
lower bound for approximate atmospheric and solar results with an
extrapolation to the first generation mass. The unification scale
marks the rough range implied by the see-saw mechanism.}
\label{fig:masses}
\end{figure}

Other chapters in this volume deal with the particle theory situation,
so we make only several general remarks here, more from the
experimentalists' phenomenological viewpoint.  Figure
\ref{fig:masses}, shows the masses of the fundamental fermions in
three generations on a logarithmic scale in mass.  Dramatically, one
sees that if the neutrino masses are near the lower bounds (that is
at the presumed mass differences from present atmospheric and solar
results), they lie 10-15 orders of magnitude below the other
fundamental fermions (charged quarks and leptons). Graphically one
notes the spacing between the neutrino masses and the charged
fermion masses is just about the same as the distance (on the log
scale) to the unification scale.  This is a pictorial representation
of the see-saw mechanism, as noted more than ten years
ago\cite{osc_hyp}.  This points up the task for grand unification
model building, and highlights the deep link between neutrino masses
and nucleon decay\cite{Mohapatra}.

        \subsection{Future Muon Neutrino Experiments}

The physics community seems to have rapidly accepted, with caution,
the seeming inevitability of neutrino mass and
oscillations\cite{Kayser}.  Yet most probably the game has hardly
begun and many a subtlety may await our exploration.  However if the
LSND claims will go quietly away (after the BOONE experiment runs),
the mass and mixing picture could settle into the simple hierarchical
pattern as explored in the bi-maximal mixing scenario (or similar
versions). This highlights the importance of experiments to follow up
on the LSND results as one of the first agenda items in the current
neutrino business (see Chapter VII).

Given present indications, it would seem that the K2K\cite{K2K},
MINOS\cite{MINOS} and CERN-Gran-Sasso experiments ICANOE\cite{ICANOE}
and OPERA\cite{OPERA} should confirm the SuperK results and make the
oscillation parameters more precise.  Of course, one would really like
to see tau appearance, not just muon disappearance to be sure we are
not being misled.  There has been some debate in the community as to
what constitutes appearance.  Because of the complexity of tau final
state identification, this author would prefer to see a real tau track
recorded.  In any case plans are in progress in the US, Japan and
Europe for the obvious follow up experiments to solidify the
oscillation scenario and refine the parameters, which experiments
should take less than a decade.  At this time it appears that only
OPERA has the opportunity to record physical tau lepton tracks in
emulsions, though the other experiments (including SuperK) may be able
to detect tau kinematic signatures statistically.

More interesting for the long range physics is filling in the MNS
matrix (lepton equivalent of quark CKM matrix) for
neutrinos\cite{MNS}.  This is not an easy business.  The atmospheric
neutrino measurements really are only defining, at best, three of the
nine elements!  Solar neutrinos get us another, perhaps a constraint
on two.  Measuring the tau related components directly seems pretty
hopeless.  Of course if we can assume the matrix to be unitary and
real we would be in good shape as there are then only three
independent parameters (plus the masses).  But we do not know this,
and if there exist CP violations we then have a total of three angles
and one phase plus possibly another phase for the case of Majorana
neutrinos, but which phase is only measurable in special circumstances
such as neutrino-less double-beta-decay.  If there are more (heavy or
sterile) neutrinos, then things could be much more complicated (as the
3 by 3 sub-matrix will not be unitary). By analogy with the quarks
(where the 3 by 3 CKM matrix with small mixing angles and one CP
violating phase suffices), perhaps we should not worry too much,
except for lack of any guidance whatsoever from theory.  CP violation
is only very weakly constrained experimentally in the neutrino sector
at present, so we could be in for big surprises, and given the
neutrino connection with cosmology and baryogenesis, one should indeed
be suspicious, I believe.

As a whole, the particle physics community is just beginning to
examine the newly illuminated possibilities in the neutrino sector.
At the moment it appears as though muon colliders may provide our best
route for next generation explorations in this
realm\cite{mumuloi,nufac}.  The serendipitous realization that muon
storage rings can provide neutrino beams more intense by about 4
orders of magnitude than previous artificial neutrino sources, along
with the potential, via polarization of these beams, to make nearly
pure beams of muon neutrinos or anti-neutrinos, with a relatively
narrow energy spread, allows one to dream of experiments not
heretofore thought possible.  The prospects for this endeavor are just
emerging at the time of this writing and the dialog between physics
possibilities and machine technical capabilities is in much ferment.
The present excitement about this endeavor, probably at least ten
years from realization, promises much evolution of thinking about
critical tests and experiment planning.  Aside from simple checking of
previous results, precision measurement of oscillation parameters, and
filling out of the MNS matrix, the major goal in this authors mind is
exploration for CP and CPT violation in the neutrino sector, not
possible by any other means yet conceived.  Of course if $\theta_{13}$
turns out to be zero, which hopefully we will know in a few years,
then CP violation is not be possible. Credibly motivated CPT violation
can still be a possibility however\cite{Barger_00}.  

Table \ref{tab:compare} shows a rough comparison of the physics
capabilities and reach of SuperK and the various present and future
long-base-line experiments.  This table was generated by an informal
working group at the 2000 Aspen ``Neutrinos with Mass'' Workshop.
Although the details may annoy proponents of some projects (and
please others), the purpose was not to evaluate specific
experiments, but to try and gauge the strengths of the various
approaches as we move forward in the quest to complete the MNS
matrix and explore for CP and CPT violations in the neutrino sector.
We had inadequate information about OPERA and the proposed 50 GeV
proton driven neutrino beam in Japan (JHF to SuperK), so they may
not be fairly represented.

The table assumes that the $\Delta m^2 = 0.003~eV^2$ and 
$\sin^2 (2\theta) \simeq 1.0$ with dominantly $\nu_\mu
\leftrightarrow \nu_{\tau}$ mixing, large mixing angle MSW solar
neutrino oscillations and no LSND/BooNE indication of oscillations.
We also assumed a $20-50~GeV$ neutrino factory and high quality
neutrino detector at a few thousand kilometers distance.

There was considerable discussion of whether the new experiments
will or will not be able to see at least one dip in the detected
neutrino spectrum, unambiguously discriminating oscillations from
disappearance (as in decay).  My conclusion is that such a detection
will be very difficult for all experiments with traditional neutrino
beams if the $\Delta m^2$ is at the low end of the allowed region.
ICANOE claims to be able to detect oscillations in the cosmic ray
beam, a consequence of their excellent resolution and the larger
range of $L/E$ values available from the atmospheric neutrinos than
from a fixed distance long-base-line beam.

I think what we all learned from this exercise is that there really
is a great gulf between what we can accomplish with a neutrino
factory and anything prior to that, even with more powerful
traditional neutrino beams.

\begin{table}
\caption{Comparison of SuperK and long-base-line neutrino experiments
in terms of physics capabilities and reach.   See text for
explanation.}
\label{tab:compare}
\begin{center}
\begin{tabular}{l c c c c c c c}
Experiment:         & SuperK    & K2K       & MINOS     & ICANOE    &   OPERA   & JHF2K     & $\nu$ Fac \\
                    &(atm $\nu$)&           &           &           &           &           &          \\
\hline
$d(ln(\Delta m^2))$ &    50\%   &    20\%   &     10\%  &    10\%   &    20\%   &    10\%    &     2\% \\
$d(\sin^2 2\theta)$ &     5\%   &     5\%   &      5\%  &     5\%   &      ?    &     4\%    &     2\% \\
See Oscillations?   & $\times$  & $\times$  & $\surd$(?)&$\times$($atm\surd$)& ? & $\times$  &  $\surd$ \\ 
$\tau$ appear (kink)& $\times$  & $\times$  & $\times$  & $\times$  & $\surd$   &  $\times$  &  $\surd$ \\
$\tau$ appear (kinem)& $\surd$? & $\times$  & 2$\surd$  & 3$\surd$  & 4$\surd$  &  $\times$  & 4$\surd$ \\
$\sin^2 2\theta_{13}$ limit & 0.1 & 0.03? & 0.03 & 0.015  & (ident e?) & 0.03  & $1-3\times 10^{-3}$ \\
$d(\nu_s / n_{\tau})$&   20\%   & $\times$  &    5\%?   &      5\%  &     ?     &  $\times$  &    1\% \\
Elim Decay  Models? & $\times$(? NC/CC)&$\times$&$\surd$(NC/CC)&$\times$(atm$\surd$)&?&$\times$&$\surd$ \\
Sign of $\Delta m^2$& $\times$  & $\times$  &  $\times$ & $\times$  & $\times$  &   $\times$ &  $\surd$ \\
$\nu_e\rightarrow\nu_{\tau}$& $\times$  & $\times$  &  $\times$ & $\times$  & $\times$  &   $\times$ &  $\surd$ \\
CP violation tests  & $\times$  & $\times$  &  $\times$ & $\times$  & $\times$  &   $\times$ &  $\surd$ \\  
CPT violation tests & $\times$  & $\times$(?)& $\surd$  &  $\surd$  & $\times$  &$\times$(?) &  $\surd$ \\
\end{tabular}
\end{center}
\end{table}

Measuring absolute neutrino mass remains a frustrating problem, which
will not be resolved in the near future it seems.  While pushing to
lower mass limits with Tritium beta decay experiments will apparently
not be able to reach below $0.1~eV$, there is some hope from CMBR
measurements\cite{Primak}, there is a long shot via the ``Weiler
process''\cite{Weiler_99}, and optimistically neutrino-less
double-beta-decay experiments may eventually reach $0.01$ or even
$0.001~eV$.

At high energies, explorations for cosmic neutrinos may be carried out
in deep arrays in the ocean and under ice\cite{Learned_00}.  While the
main goal of these attempts at high energy neutrino astronomy will be
aimed at astrophysics, with a high-energy-threshold-detector capable
of registering neutrinos in the $PeV$ range, it may be that such
instruments will be able to directly detect tau neutrinos (via the
``double bang'' signature\cite{Learned_95}), and even determine the
neutrino flavor mix, to the benefit of both particle physics and
astrophysics.

A next generation (megaton) scale nucleon decay instrument to probe
lifetimes to $10^{35}~years$ would do wonders for advancing neutrino
physics as well.  Simply building a larger version of SuperK will
not suffice because of the need for greater resolution as well as
size.  The only candidate I see to go beyond SuperK is something
like the AQUA-Rich style of imaging water Cherenkov
detector\cite{Ypsilantis,Learned_98}.  An attractive alternative
which need not be so massive to get to $10^{35}$ years in kaon modes
of nucleon decay, might be a $50-70~kiloton$ liquid argon detector of
the ICARUS style.  Perhaps such a detector can be realized in
concert with a long baseline beam from a neutrino factory.

From the foregoing it should be apparent that we have entered a new
era in elementary particle physics, and that one can expect a long and
interesting exploration into neutrino mass and mixing now that the
door has been opened.

\section*{Acknowledgments}

As noted earlier, the Super-Kamiokande Collaboration deserves the
credit for most of the work reported herein, and any errors in
interpretation are those of the author.  Thanks to Sandip Pakvasa
for many discussions and much help.  Thanks also to Tony Mann, from
whose excellent summary\cite{Mann_99} I drew heavily. And finally
thanks to the Aspen Center for Physics, and the year 2000
``Neutrinos with Mass'' confreres for many lively neutrino
discussions.

\input{references}
\input{appendix}

\end{document}

%% file: appendix.tex
\appendix
\section*{Appendix: Super-Kamiokande Collaboration, 6/00}





\begin{description}

\item{{\bf Boston U. } M.~Earl, A.~Habig, E.~Kearns, M.D.~Messier, 
K.~Scholberg, J.L.~Stone,L.R.~Sulak, C.W.~Walter}

\item{{\bf Brookhaven Nat. Lab. } M.~Goldhaber}

\item{{\bf U. of Cal., Irvine }
T.~Barszczak, D.~Casper, W.~Gajewski, W.R.~Kropp, S.~Mine, 
L.R.~Price, M.~Smy, H.W.~Sobel, M.R.~Vagins}

\item{{\bf Cal. State U., Dominguez Hills } K.S.~Ganezer, W.E.~Keig}

\item{{\bf George Mason U. } R.W.~Ellsworth}

\item{{\bf Gifu U.} S.~Tasaka}

\item{{\bf U. of Hawaii, Manoa}
A.~Kibayashi, J.G.~Learned, S.~Matsuno, D.~Takemori}

\item{{\bf Inst. of Part. and Nuc. Stud., KEK}
Y.~Hayato, T.~Ishii, T.~Kobayashi, K.~Nakamura, Y.~Oyama, 
A.~Sakai, M.~Sakuda, O.~Sasaki}

\item{{\bf Kobe U.} S.~Echigo, M.~Kohama, A.T.~Suzuki}

\item{{\bf Kyoto U.} T.~Inagaki, K.~Nishikawa}

\item{{\bf Los Alamos Nat. Lab. } T.J.~Haines}

\item{{\bf Louisiana State U. } E.~Blaufuss, B.K.~Kim, R.~Sanford,
R.~Svoboda}

\item{{\bf U. of Maryland } M.L.~Chen, J.A.~Goodman, 
G.~Guillian, G.W.~Sullivan}

\item{{\bf State U. of N.Y., Stony Brook } 
J.~Hill, C.K.~Jung, K.~Martens, M.~Malek, C.~Mauger, 
C.~McGrew, E.~Sharkey, B.~Viren, C.~Yanagisawa}

\item{{\bf Niigata U.}
M.~Kirisawa, S.~Inaba, C.~Mitsuda, K.~Miyano, H.~Okazawa, 
C.~Saji, M.~Takahashi, M.~Takahata}

\item{{\bf Osaka U.} Y.~Nagashima, K.~Nitta, M.~Takita, M.~Yoshida}

\item{{\bf Seoul Nat. U.} S.B.~Kim}

\item{{\bf Tohoku U.} M.~Etoh, Y.~Gando, T.~Hasegawa,
K.~Inoue, K.~Ishihara, T.~Maruyama, J.~Shirai, A.~Suzuki}

\item{{\bf U. of Tokyo} M.~Koshiba}

\item{{\bf Inst. for Cos. Ray Res., U. of Tokyo}
S.~Fukuda, Y.~Fukuda, M.~Ishitsuka, Y.~Itow, T.~Kajita, J.~Kameda,
K.~Kaneyuki, K.~Kobayashi, Y.~Kobayashi, Y.~Koshio, M.~Miura, 
S.~Moriyama, M.~Nakahata, S.~Nakayama, Y.~Obayashi, A.~Okada, 
K.~Okumura, N.~Sakurai, M.~Shiozawa, Y.~Suzuki, H.~Takeuchi, 
Y.~Takeuchi, T.~Toshito, Y.~Totsuka (Spokesman), S.~Yamada}

\item{{\bf Tokai U.} Y.~Hatakeyama, Y.~Ichikawa, 
M.~Koike, K.~Nishijima}

\item{{\bf Tokyo Inst. for Tech.}
H.~Fujiyasu, H.~Ishino, M.~Morii, Y.~Watanabe}

\item{{\bf Warsaw U.} U.~Golebiewska, D.~Kielczewska}

\item{{\bf U. of Washington, Seattle } S.C.~Boyd, A.L.~Stachyra,
R.J.~Wilkes, K.K.~Young}

\end{description}

%% file: jgl.bbl
\begin{thebibliography}{99}
\bibitem{pull} Nonetheless, this author (personal opinion, not
collaboration) suspects that the process pulls the minima slightly
towards lower values of $\Delta m^2$.  The reasoning is that the
introduced parameters un-weight the effect of $R$ on the fitting,
which pulls upwards, while the shape pulls downwards.  The author's
bet remains that $\Delta m^2$ settles at around $5 \times
10^{-3}~eV^2$, whereas the official fits give $2.5~to~3.5$.
\end{thebibliography}
